\documentclass[%
 reprint,
 amsmath,amssymb,
 aps,
]{revtex4-2}

\usepackage{graphicx}% Include figure files
\usepackage{dcolumn}% Align table columns on decimal point
\usepackage{bm}% bold math
\usepackage{booktabs} 
\usepackage[T1]{fontenc}

\usepackage{siunitx}
\usepackage{hyperref}% add hypertext capabilities
 \hypersetup{ 
     colorlinks=true, 
     linkcolor=blue, 
     filecolor=blue, 
     citecolor = blue,       
     urlcolor=black, 
     } 
\usepackage[capitalise]{cleveref}

\begin{document}

\preprint{APS/123-QED}

\title{Industrial 300$\,$mm wafer processed spin qubits in natural silicon/silicon-germanium}% Force line breaks with \\
%#\thanks{A footnote to the article title}%

\author{Thomas Koch\textsuperscript{1}}
 \email{T.Koch@kit.edu}
\author{Clement Godfrin\textsuperscript{2}}
\author{Viktor Adam\textsuperscript{3}}
\author{Julian Ferrero\textsuperscript{1}}
\author{Daniel Schroller\textsuperscript{1}}
\author{Noah Glaeser\textsuperscript{1}}
\author{Stefan Kubicek\textsuperscript{2}}
\author{Ruoyu Li\textsuperscript{2}}
\author{Roger Loo\textsuperscript{2,4}}
\author{Shana Massar\textsuperscript{2}}
\author{George Simion\textsuperscript{2}}
\author{Danny Wan\textsuperscript{2}}
\author{Kristiaan De Greve\textsuperscript{2,5}}
\author{Wolfgang Wernsdorfer\textsuperscript{1,3}}
 \email{Wolfgang.Wernsdorfer@kit.edu}

\affiliation{ 
\textsuperscript{1}Physikalisches Institut, Karlsruhe Institute of Technology (KIT), Karlsruhe, Germany
}%
\affiliation{ 
\textsuperscript{2}Interuniversity Microelectronics Centre (imec), Leuven, Belgium
}%
\affiliation{ 
\textsuperscript{3}Institute for Quantum Materials and Technologies, Karlsruhe, Germany
}%
\affiliation{
\textsuperscript{4}Department of Solid-State Sciences, Ghent University, Ghent, Belgium
}
\affiliation{ 
\textsuperscript{5}Department of Electrical Engineering - ESAT-MNS, KU Leuven, Leuven, Belgium
}%

\date{\today}% It is always \today, today,
             %  but any date may be explicitly specified

\begin{abstract}
The realisation of an universal quantum computer will require the operation of thousands to millions of qubits. The possibility of using existing industrial semiconductor fabrication techniques and infrastructure for up-scaling and reproducibility makes silicon based spin qubits one of the most promising platforms to achieve this goal. The implementation of the up to now largest semiconductor based quantum processor was realized in a silicon/silicon-germanium heterostructure known for its low charge noise, long qubit coherence times and fast driving speeds, but the high structural complexity creates challenges for industrial implementations. Here we demonstrate quantum dots hosted in a natural Si/SiGe heterostructure fully fabricated by an industrial \SI{300}{\milli\metre} semiconductor wafer process line from heterostructure growth to Co micromagnet monolithic integration. We report charge noise values below \SI{2}{\micro\electronvolt / \sqrt{Hz}}, spin relaxation times exceeding \SI{1}{s}, and coherence times $T_2^*$ and $T_2^H$ of \SI{1}{\micro\second} and \SI{50}{\micro\second} respectively, for quantum wells grown using natural silicon. Further, we achieve Rabi frequencies up to \SI{5}{MHz} and single qubit gate fidelities above \SI{99}{\percent}. In addition to scalability, the high reproducibility of the \SI{300}{\milli\metre} processes enables the deterministic study of qubit metric dependencies on process parameters, which is essential for optimising qubit quality.
\end{abstract}

%\keywords{Suggested keywords}%Use showkeys class option if keyword
                              %display desired
\maketitle

%\tableofcontents
\section{Introduction}\label{sec1}

A quantum computer capable of solving relevant problems requires a large amount of interconnected qubits paired with high-fidelity gate operations. Semiconductor based spin qubits are one of the most promising platforms in order to achieve this goal \cite{loss_quantum_1998,vandersypen_interfacing_2017}. Single and two qubit gate fidelities exceeding the error correction threshold have been demonstrated \cite{yoneda_quantum-dot_2018, xue_benchmarking_2019, huang_fidelity_2019}. High-fidelity operations at elevated temperatures increase the available cooling power drastically and ease the combination with classical control electronics \cite{ono_high-temperature_2019, undseth_hotter_2023, huang_high-fidelity_2024}. Long range coupling can be implemented by superconducting resonators or electron shuttles \cite{seidler_conveyor-mode_2022, dijkema_two-qubit_2023, de_smet_high-fidelity_2024}. The realisation of the up to date largest semiconductor based quantum processor consists of six qubits hosted in quantum dots formed in an isotopically purified $^{28}$Si quantum well strained between silicon-germanium layers \cite{philips_universal_2022}. The Si/SiGe heterostructure aims to decouple the quantum well from the defective semiconductor oxide interface, which is known as the main contribution for charge noise inside the sample \cite{shehata_modeling_2023, fleetwood_effects_1993, dutta_low-frequency_1981}. Paired with micromagnets that enable qubit manipulation by electric dipole spin resonance (EDSR), Rabi frequencies of several MHz and gate fidelities exceeding the error correction threshold  were demonstrated \cite{yoneda_quantum-dot_2018}. Meanwhile, industrial CMOS fabrication technologies have advanced to the point where they can produce classical processors containing up to trillions of MOSFETs \cite{rocki_fast_2020}. Utilizing these fabrication techniques in the production of semiconducting qubits has the potential to enhance reproducibility and uniformity, as well as simplify the optimisation and fine-tuning of qubit parameters. These are all crucial steps towards the realisation of a universal quantum computer. Qubits fabricated with industrial semiconductor manufacturing technologies have been demonstrated recently for Si-MOS \cite{maurand_cmos_2016, zwerver_qubits_2022, klemt_electrical_2023} and Si/SiGe heterostructures \cite{neyens_probing_2024} and report quantum dot yields above \SI{99}{\percent}. However, the qubit control modules, specifically the micromagnets used for fast EDSR control, have not yet been fully integrated into the process flow. In this work, we demonstrate the operation of EDSR qubits in a natural Si/SiGe heterostructure, fully manufactured by an industrial \SI{300}{mm} semiconductor wafer process line, with a cobalt micromagnet control module incorporated through monolithic integration.

\section{Results}\label{sec2}

\begin{figure*}[t]%
\centering
\includegraphics[width=1\textwidth]{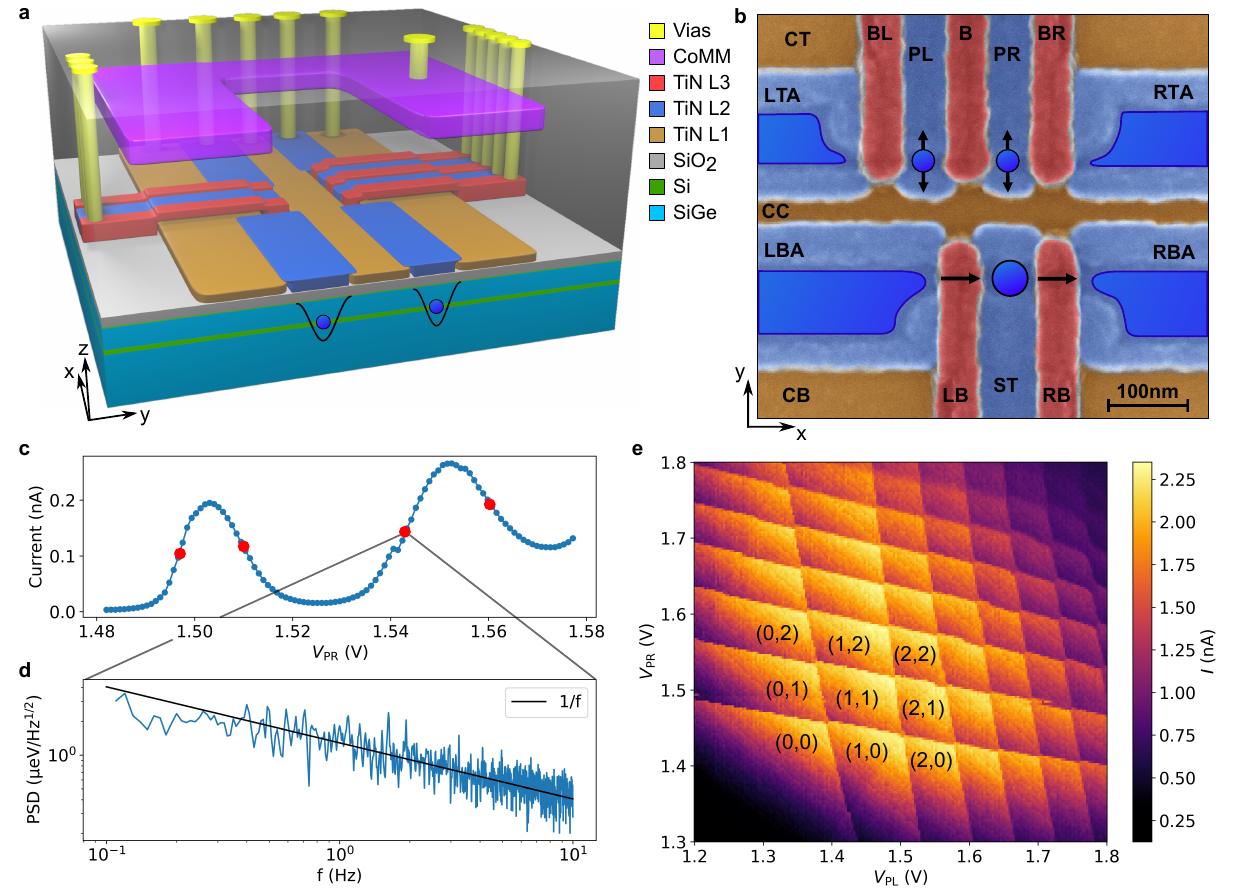}
\caption{\textbf{a} 3D model of the device. 3-layer TiN gate architecture (orange, blue, red) on the SiGe heterostructure (cyan and green). SiO$_2$ (grey) insulates the different layers from each other. The electron gas is formed in the buried Si layer. The CoMM (violet) is encapsulated by a SiO$_2$ passivation layer, and all structures are connected by vias (yellow). \textbf{b} False coloured scanning electron microscopy image with included gate labeling and schematic display of later formed electron reservoirs and quantum dot positions. The CC gate will be pulsed with RF microwaves and the arrows indicate the direction of the oscillating quantum dot positions and the current path of the sensor dot. \textbf{c} Coulomb oscillations of the right qubit dot, formed below gate PR. Red dots indicate the 4 positions used to extract values for charge noise. \textbf{d} Calculated power spectral density of a \SI{5}{min} timetrace taken at the indicated working point. A mean \SI{1}{Hz} value of \SI{1.36(0.07)}{\micro\electronvolt / \sqrt{Hz}} is extracted. The black line follows a 1/f proportionality. \textbf{e} Charge stability diagram of the two coupled qubit quantum dots, measured via charge sensing by the sensor dot. Both qubit plunger gates are swept and every visible line corresponds to a change of electron number in the respective dots, starting with a (0,0) occupation in the lower left.}\label{fig1}
\end{figure*}

The natSi/SiGe heterostructures are grown on a \SI{300}{\milli\metre} silicon wafer. From bottom to top, the heterostructure comprises a \SI{8}{\micro\metre} strained relaxed buffer (SRB) finished by an unstrained \SI{1}{\micro\metre} Si$_{0.75}$Ge$_{0.25}$ layer of constant concentration. This is followed by chemical vapor deposition (CVD) of a tensile-strained natSi quantum well of \SI{9}{\nano\metre} and a Si$_{0.75}$Ge$_{0.25}$ buffer layer of \SI{40}{nm} thickness. To avoid oxidation of the SiGe buffer, a \SI{2}{nm} thick silicon cap is grown on top of the heterostructure. Ohmic contacts to the natSi quantum well are formed by phosphorus implants. Three overlapping gate layers made of titanium nitride (TiN) and insulated by silicon oxide (SiO$_2$) (From bottom to top: \SI{8}{nm} SiO$_2$, \SI{30}{nm} TiN, \SI{5}{nm} SiO$_2$, \SI{20}{nm} TiN, \SI{5}{nm} SiO$_2$, \SI{20}{nm} TiN) are used for the formation and manipulation of the quantum dots. A cobalt micromagnet (CoMM) module is made in line using a damascene process (for more details on the CoMM Supplementary Information \cref{Supfig3}). A SiO$_2$ passivation layer is deposited on top of the stack. Vias are etched in this passivation layer to connect the ohmics, gate layers and the CoMM (see \cref{fig1}\textbf{a}). The gate layout of the device is shown in the false colored scanning electron microscopy (SEM) picture in \cref{fig1}\textbf{b}. The architecture is designed to enable the formation of two quantum dots in the single electron regime, later used as qubits, and an adjacent single electron transistor (SET) for readout. The qubit dots are centered below the gap of a C-shaped CoMM, which creates an inhomogeneous external magnetic field at the qubit positions. By applying a microwave signal at the central screening gate (CC), the position of the qubit dot will oscillate, effectively creating an AC magnetic field with the frequency of the applied microwave signal \cite{pioro-ladriere_electrically_2008}. \cref{fig1}\textbf{c} shows the measured tunnel current through the right qubit quantum dot over varying voltages applied at the plunger gate PR. Coulomb oscillations of the measured current become visible. It was shown that the charge noise affecting a quantum dot can strongly differ for different working points \cite{elsayed_low_2024}. In order to extract a value for the charge noise, we measure the power spectral density at four different working points individually. \cref{fig1}\textbf{d} shows the power spectral density at the marked working point, which follows a 1/$f$ proportionality often observed for charge noise in semiconducting quantum dot devices \cite{paladino_1_2014, ferrero_noise_2024}. The mean charge noise value at \SI{1}{Hz} from the four working points results to \SI{1.36(0.07)}{\micro\electronvolt / \sqrt{Hz}} (for more details see Methods \ref{ChargeNoise}). \cref{fig1}\textbf{e} shows the measured sensor current while varying both qubit plunger gate voltages. The sensor current jumps every time the electron occupation of the qubit dots changes. This charge stability diagram enables precise control of the charge state of the qubits, starting from zero electron occupation in the bottom left, up to multiple electrons in both dots in the upper right. 

\begin{figure}[t]%
\centering
\includegraphics[width=0.5\textwidth]{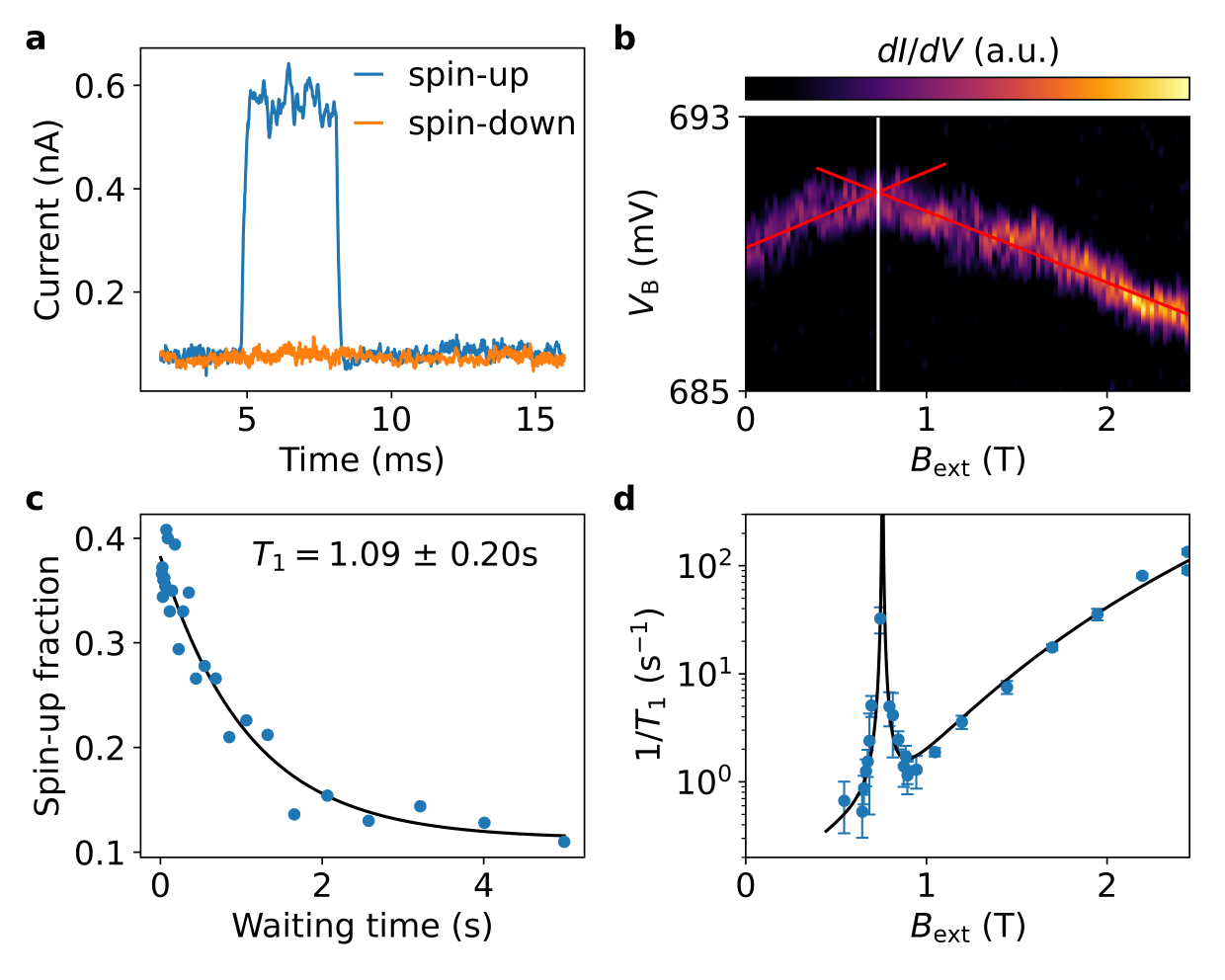}
\caption{\textbf{a} Measured real time SET current for a spin-up electron (blue) compared to a spin-down electron (orange). \textbf{b} Numerical derivation of the SET current tracking the 1$\rightarrow$2 electron transition of the qubit dot for external magnetic fields up to \SI{2.5}{T}. The barrier gate B is used as plunger gate because of the lower lever arm on the dot potential. The change in slope of the tracked transition occurs at the external field, which is equal to the valley splitting. Linear fits to the range from \SI{0}{T} to \SI{0.4}{T} and \SI{1.3}{T} to \SI{2.5}{T} result in a crossing at \SI{0.73}{T} which is equal to a valley splitting of \SI{84.5}{\micro\electronvolt}. \textbf{c} Elzerman pulse sequence for varying waiting times of the loading pulse at an external magnetic field of \SI{0.66}{T}. The exponential decay corresponds to a spin-relaxation time $T_1 = \SI{1.09(0.20)}{s}$. \textbf{d} Spin relaxation rate $1/T_1$ over external magnetic field. The black curve shows a fit combining Johnson noise and spin-phonon interaction diverging at the value corresponding to the valley splitting energy. The fit diverges at an external field of \SI{0.76}{T} resulting in a valley splitting of \SI{88.0}{\micro\electronvolt}.}\label{fig2}
\end{figure}

By applying an external magnetic field in x-direction, the ground state degeneracy of the electron spin state is lifted via the Zeeman effect. By tuning the quantum dot potential to an energy level between the Zeeman-split spin-down and spin-up states of the electron, a spin selective readout becomes possible. This is demonstrated in \cref{fig2}\textbf{a}, where the measured sensor current jumps between two distinct values in case of a spin-up event, or stays constant for an electron in the spin-down state. The spin relaxation time can be measured by a three stage pulse scheme \cite{elzerman_single-shot_2004}, incorporating the spin selective readout mentioned before. In the first stage the quantum dot is emptied. In the second stage, the quantum dot potential is pulsed deep below the chemical potential of the reservoir to enable the possibility for a tunneling electron to occupy either the ground or excited spin state. And in the third stage the quantum dot is pulsed to a level where the reservoir lies between the ground and excited spin state, to perform the spin selective readout. By varying the duration of the loading pulse, we increase the available time for a spin relaxation process to occur, which is dominated by spin-phonon interaction \cite{huang_spin_2014, hollmann_large_2020, borjans_single-spin_2019}. \cref{fig2}\textbf{c} shows the measured spin-up fraction of 500 single shot readouts performed for different waiting times in the loading stage, at an external magnetic field of \SI{660}{mT}, which is later used for qubit manipulations at resonance frequencies of $\approx \SI{18.5}{\giga\hertz}$. The spin-up fraction over waiting time is proportional to $\exp{(-t/T_1)}$, and the fit results in a spin relaxation time of $T_1 = \SI{1.09(0.20)}{s}$. 

Another important quantity for Si/SiGe heterostructures is the valley splitting energy of the two lowest valley states \cite{hollmann_large_2020, burkard_semiconductor_2023, zwanenburg_silicon_2013}. The conduction band minimum of silicon is normally sixfold degenerate. Straining the Si quantum well between the two SiGe layers splits the degeneracy in two energetically favourable valley states and four excited valley states. The degeneracy of the two lowest valley states gets further lifted by electrical confinement of the electron wave function, and strongly depends on sample geometry and local inhomogeneities \cite{hosseinkhani_electromagnetic_2020, volmer_mapping_2024}. To measure the valley splitting at the position of the qubit quantum dot we performed two different measurements. In \cref{fig2}\textbf{b}, the plunger gate voltage of the qubit quantum dot is varied for different total external magnetic fields (externally applied field plus CoMM contribution in saturation, see \cref{Supfig3} in Supplementary Information for details). The colour axis represents the numerical derivation of the measured current through the sensor dot. The gate voltage window required for the transition from 1 to 2 electrons inside the dot is shown. The exact position of the 1$\rightarrow$2 electron transition shows a change of slope at an external field strength equal to the valley splitting energy. To extract the two slopes, we fit the data in the 0 to \SI{0.4}{T} interval and the 1 to \SI{2.5}{T} interval separately and receive a crossing of both lines at \SI{0.73(8)}{T}, which corresponds to \SI{85(9)}{\micro\electronvolt}. Another method is by measuring the spin relaxation time over magnetic field. Here we observe a drastic decrease of the $T_1$ time, visible as a peak of the spin relaxation rate $1/T_1$, at the position where the Zeeman energy is again equal to the valley splitting energy. We fit a rate equation that accounts for spin-valley mixing and intra-valley spin-orbit interaction in combination with phonon noise and Johnson noise \cite{hollmann_large_2020,yang_spin-valley_2013, petit_spin_2018}. The fit gives a peak position of \SI{0.76(1)}{T} which results in a valley splitting of \SI{88(1)}{\micro\electronvolt}. Both extracted values for the valley splitting are similar and correspond to a frequency above \SI{20}{GHz}.

\begin{figure}[t]%
\centering
\includegraphics[width=0.5\textwidth]{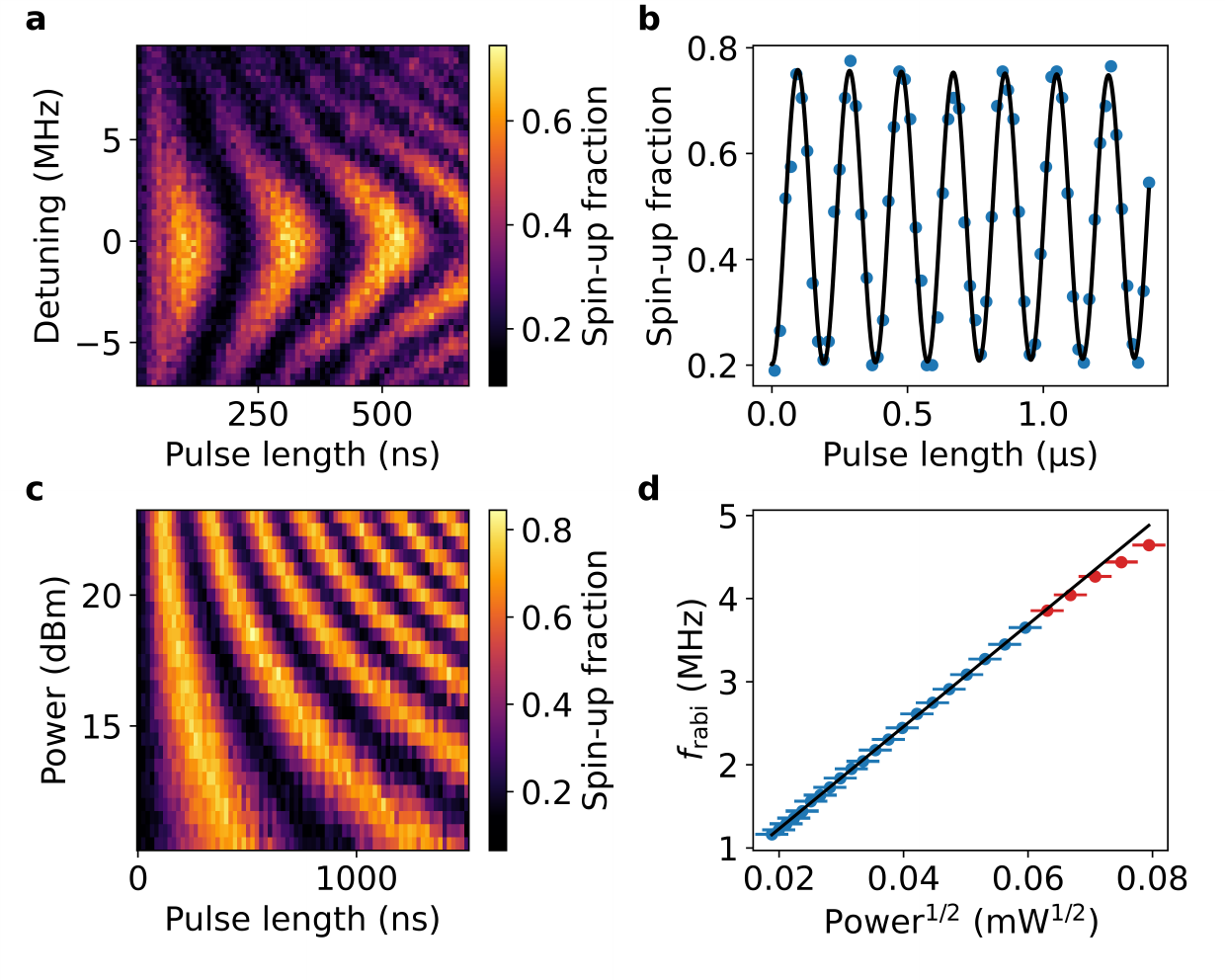}
\caption{\textbf{a} Rabi chevron pattern at \SI{23}{dBm} RF output power of the microwave source. \textbf{b} Rabi measurement on resonance with \SI{24}{dBm} output power up to a pulse length of \SI{1.5}{\micro\second}. A Rabi frequency of \SI{5.2}{\mega\hertz} is extracted by fitting a sine function to the data. No dampening of the frequency amplitude is visible up to this pulse length. \textbf{c} Same measurement as in \textbf{b} for varying RF output powers. \textbf{d} Extracted Rabi frequencies for all used powers of the measurement shown in \textbf{c} over the applied power converted to $\sqrt{\mathrm{mW}}$. The black line corresponds to a linear fit excluding the 5 data points of highest applied power (marked in red).}\label{fig3}
\end{figure}

For coherent spin manipulation we report measurements for three different devices. Device A with a CoMM gap size of \SI{300}{nm}, and devices B and C with a CoMM gap of \SI{650}{nm}. Additionally, devices A and C origin from the same \SI{300}{mm} wafer, while device B stems from a separate one (measurements of devices B and C in Supplementary Information). For the measurements displayed in \cref{fig3}, every extracted spin-up fraction is the result of 200 single shot readouts. \cref{fig3}\textbf{b} shows the spin-up fraction over manipulation pulse duration at qubit resonance frequency of \SI{18.501}{\giga\hertz}. Typical Rabi oscillations become visible. Fitting the oscillations gives a Rabi frequency of \SI{5.2}{\mega\hertz} for an applied RF microwave power of \SI{24}{dBm}. \cref{fig3}\textbf{a} shows a Chevron pattern, resulting from several Rabi measurements for different frequency detunings at \SI{23}{dBm} output power. The whole measurement was taken over a span of \SI{7}{h} and also demonstrates the long term stability of the qubit frequency and readout working point. \cref{fig3}\textbf{c} displays the spin up fractions for different microwave source output powers in dBm over pulse duration and \cref{fig3}\textbf{d} the extracted Rabi frequency for the respective square root of the power (10 to \SI{23}{dBm} subtracted by \SI{45}{dB} inline attenuation and converted to mW). The Rabi frequencies follow a clear linear behaviour over $\sqrt{\mathrm{power}}$ up to values above $\approx \SI{0.07}{\sqrt{\milli\watt}}$, where a saturation starts to set in. The black line corresponds to a linear fit of the data excluding the 5 resulting Rabi frequencies for the highest powers. Saturation is a sign of a different qubit drive regime, which could be caused by either a lower magnetic field gradient for higher RF amplitudes or a strong quantum dot confinement in the x-direction, dampening larger displacements of the quantum dot position.

\begin{figure}[b]%
\centering
\includegraphics[width=0.5\textwidth]{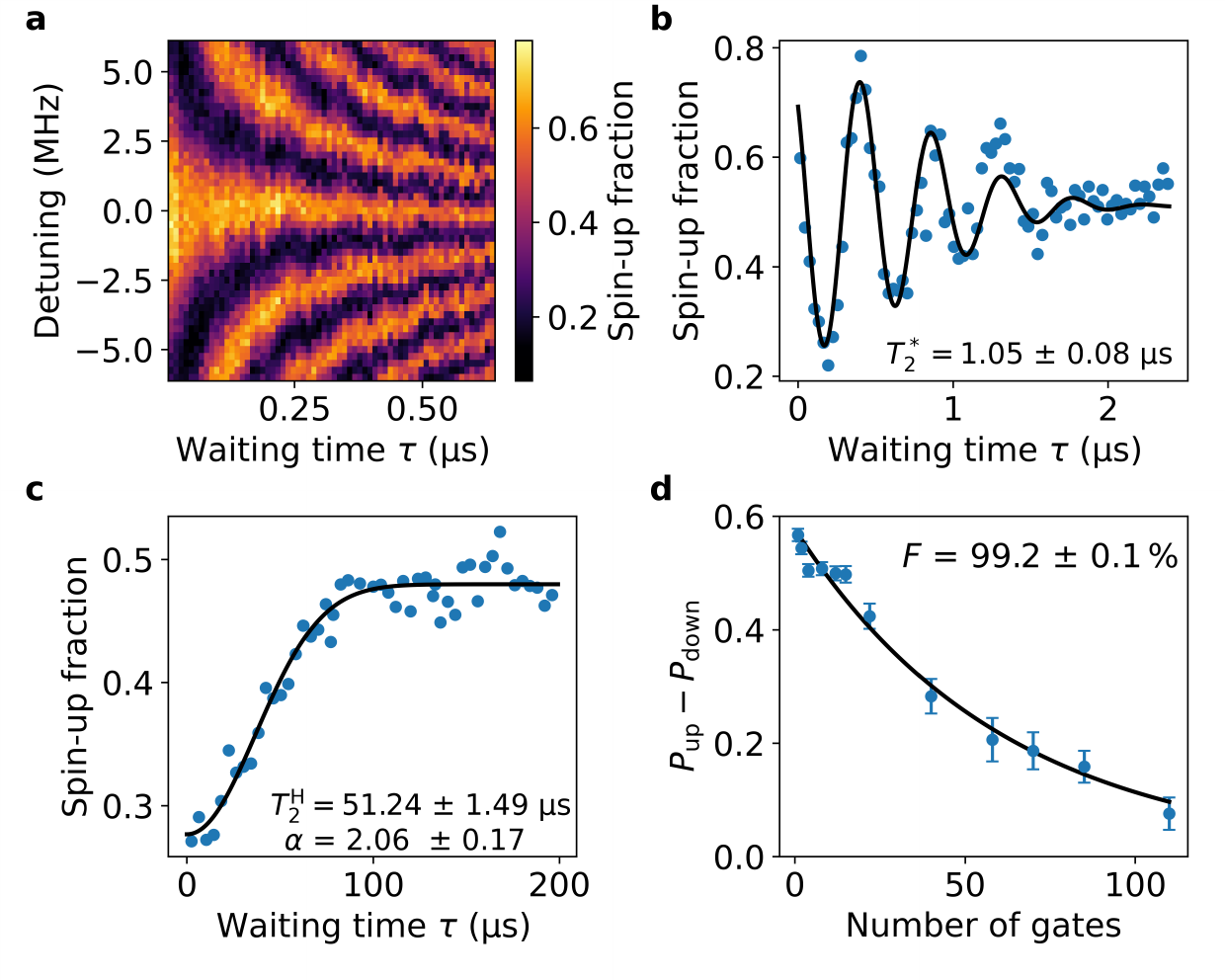}
\caption{\textbf{a} Measured spin-up fraction after a Ramsey pulse sequence for varying frequency detuning and waiting time $\tau$ between the two $\pi/2$ pulses, measured over the course of $\approx \SI{6}{h}$. \textbf{b} Ramsey measurement for a fixed detuning of \SI{2}{MHz}. The black curve corresponds to a gaussian damped sine and gives a $T_2^*$ value of \SI{1.05(0.08)}{\micro\second}. \textbf{c} Spin-echo measurement fitted with $\exp(-(t/T_2^{\mathrm{H}})^\alpha$), yielding $T_2^{\mathrm{H}} = \SI{51.2(1.5)}{\micro\second}$ with $\alpha = 2.06 \pm 0.17$. \textbf{d} Difference of spin-up and spin-down fraction depending on the length of a randomized gate sequence. The black curve corresponds to a fit of the form $A\cdot P^n$ which yields the primitive gate fidelity $F = 1 - (1-P)/2 = \SI{99.2(1)}{\percent}$.}\label{fig4}
\end{figure}

\cref{fig4}\textbf{a} shows a 2D Ramsey experiment in which the waiting time $\tau$ between two $\pi /2$ pulses is varied for different frequency detunings. On resonance the spin-up fraction decays exponentially over the decoherence time $T_2^*$. When detuned, the qubit will perform a z-rotation on the Bloch sphere with the same frequency as the detuning frequency, leading to the visible pattern and the so called Ramsey fringes. \cref{fig4}\textbf{b} shows a Ramsey measurement with a fixed detuning of \SI{2}{MHz}. Fitting the oscillations with a sine multiplied by a $\exp(-(t/T_2^*)^2)$ term gives a decoherence time $T_2^*= \SI{1.05(0.08)}{\micro\second}$, similar to the reported values in other natural silicon spin qubit devices \cite{stano_review_2022} and consistent for all three measured qubit devices in this work (see Supplementary Information). \cref{fig4}\textbf{c} displays the results for a Hahn echo experiment, fitted with $\exp(-(t/T_2^\mathrm{H})^\alpha)$. The extracted $T_2^\mathrm{H}$ equals \SI{51.24(1.49)}{\micro\second} with an exponent $\alpha = 2.06 \pm 0.17$ corresponding to a gaussian damped exponential decay, observed in systems where the dominant decoherence mechanism follows a $1/f$ frequency dependence \cite{hanson_spins_2007, paladino_1_2014, mccourt_learning_2023}. The comparison between $T_2^*$ and $T_2^\mathrm{H}$ shows the frequency dependence of the dominant decoherence channel, indicating that the noise amplitudes increase to lower frequencies. Considering the abundance of nuclear spins of $^{29}$Si isotopes in natural silicon, it was demonstrated that fluctuating nuclear spins in the vicinity of the qubit are the dominant decoherence mechanism \cite{zwanenburg_silicon_2013, xue_benchmarking_2019}. Isotopic enrichment of $^{28}$Si can therefore drastically reduce this low-frequency decoherence channel and thus increase the $T_2^*$ time by orders of magnitude \cite{muhonen_storing_2014, neyens_probing_2024, philips_universal_2022}. To measure the single qubit gate fidelity, we performed randomized benchmarking experiments. For this, starting from a defined initial state, a random sequence of $n$ Clifford gates is performed. A final gate ideally brings the qubit back to a desired final state for readout. With increasing $n$ the probability to reach this desired final state will decrease due to gate errors, and finite qubit lifetimes. By fitting the deviation of the ideal final state over the length $n$ of the random sequence, we can extract the single qubit gate fidelity. To increase the robustness to readout working point deviations, we perform each random gate sequence with a final projection into the spin-up state and a final projection into the spin-down state, and subtract the resulting spin-up probabilities from each other to obtain the difference $P_\mathrm{up}-P_\mathrm{down}$ between the two final states. This difference starts at a maximum value for a small $n$ and converges to zero for large $n$, where the quantum information is completely lost. In the experiment the primitive gate set \{I, $\pm$X, $\pm$Y, $\pm$X$^2$, $\pm$Y$^2$\} is used to construct all 24 single qubit Clifford gates, where a Clifford gate consists in average of 1.875 primitive gates (for details see supplemental materials \ref{sup_Gates}). \cref{fig4}\textbf{d} shows the result of the performed randomized benchmarking. Each measured $P_\mathrm{up}-P_\mathrm{down}$ value is the result of 40 randomized gate sequences of length $n$ and is fitted with $A\cdot P^n$ from which we extract the gate fidelity $F = 1-(1-P)/2$. For the primitive gate fidelity, we extract a value of \SI{99.2(1)}{\percent} which converts to a Clifford gate fidelity of \SI{98.5(1)}{\percent}.

\section{Conclusion}\label{sec13}

In conclusion, we demonstrate the formation and operation of electric dipole spin resonance qubits in natural Si/SiGe heterostructures fully fabricated and grown in a \SI{300}{mm} industrial semiconducting wafer process line. We report a charge noise value of \SI{1.36(0.07)}{\micro\electronvolt / \sqrt{Hz}} and valley splitting energies above \SI{84}{\micro\electronvolt}. We achieve spin relaxation times $T_1$ above \SI{1}{s} and decoherence times $T_2^*$ and $T_2^\mathrm{H}$ of \SI{1}{\micro\second} and \SI{50}{\micro\second} respectively. Paired with Rabi frequencies up to \SI{5}{MHz} we achieve single qubit gate fidelities of \SI{99.2}{\percent}. Fluctuating nuclear spins of $^{29}$Si isotopes in natural silicon are known to limit qubit lifetimes especially in the low frequency regime, and isotopic purification of $^{28}$Si in the quantum well is expected to enhance especially the $T_2^*$ time by at least one order of magnitude. In addition, optimisation of the material, geometry and position of the micromagnet could further increase the manipulation speed of the qubit and improve gate fidelity. Overall, these results demonstrate the operation of state of the art spin qubit devices following industrial \SI{300}{mm} wafer process design rules. This opens the door for the optimisation of qubit metrics on a large scale and the development of much larger and more complex spin qubit architectures required for the realisation of a universal quantum computer.

\begin{acknowledgments}
We thank C. Sürgers for general advice and help with the measurement electronics and T. Cubaynes for the design of the sample printed circuit board. We further thank the German Research Foundation (DFG) for the Gottfried Wilhelm Leibniz-Award, ZVN-2020 WE 4458-5 for financial support. The work at imec was supported by the imec Industrial Affiliation Program on Quantum Computing, and the European Union’s Horizon 2020 Research and Innovation Program under grant agreement No 951852 (QLSI) 
\end{acknowledgments}

\section{Methods}\label{sec11}

\subsection{Measurement setup}\label{MeasSetup}
The samples were cooled in a Qinu Version XL dilution refrigerator, with an operating temperature of about \SI{30}{mK}. The DC lines are filtered by two RC low pass filter stages at mK with cutoff frequencies of \SI{250}{Hz} for the static gate voltages, and \SI{40}{kHz} for DC lines used for current measurements through the charge sensor and qubit plunger voltage pulses. Additionally, every line was filtered by a LFCN-80+ for mid range frequency filtering between \SI{225}{\mega\hertz} and \SI{4500}{\mega\hertz}. An electron temperature of \SI{120}{mK} is extracted from the width of the 0$\rightarrow$1 qubit charge transition. RF pulses are sent through CuNi coaxial cables with \SI{26}{dB} attenuation at \SI{18.5}{\giga\hertz} at room temperature, and an additional \SI{19}{dB} inline attenuation. The RF signal is capacitively coupled to the DC line on the printed circuit board which is wire bonded to the central screening gate (CC) of the sample. All static voltages were applied by an Adwin Pro II from Jaeger Messtechnik. The measured current through the sensor dot was converted to a voltage at room temperature by an SP983c Basel Precision Instruments I-V converter, and the voltage was digitized by an analogue input module of the Adwin Pro II. Qubit plunger voltage pulses were realised with an HDAWG from Zurich Instruments. EDSR microwave manipulation was done by controlling the in-phase (I) and quadrature (Q) channels of an SMW200A Rohde \& Schwarz vector signal generator by two HDAWG outputs.

\subsection{Charge noise measurements}\label{ChargeNoise}

The \SI{1}{Hz} charge noise value for a certain working point is extracted by taking a \SI{5}{min} long timetrace of the current data. The timetrace is converted to a power spectral density using the Welch method. Welch’s method \cite{welch_use_1967} computes the power spectral density by dividing the timetrace into overlapping segments and averaging them. Additionally, a window function is applied to the individual segments. The use of the window function averages noise in the estimated power spectrum in exchange for a reduced frequency resolution because of overlapping time windows. A Hann function was used as the window function. The resulting power spectral density is then normalized with the slope of the Coulomb oscillation at the used working point and converted to eV by multiplying with the plunger gate lever arm. The lever arm for gate PR was derived from Coulomb diamond measurements in which the source-drain bias voltage is varied over changing plunger gate voltages, and results in \SI{0.10(0.01)}{\electronvolt / \volt}. The final power spectral density is then fitted with a $a/f^b$ function and the value for \SI{1}{Hz} is extracted.

\subsection{Single shot spin selective readout}\label{Readout}

In order to differentiate between the spin-up and spin-down state of a tunneling electron, we tune the plunger gate voltage of the qubit dot so that the Fermi energy of the reservoir lies between the Zeeman-split ground state energies of the electron in the qubit quantum dot. The tunneling rate between qubit dot and reservoir is tuned to $\approx \SI{1}{kHz}$ and the standard readout time window used for the single shot measurements is \SI{15}{ms}. Depending on the sign of the current jump caused by a tunnelling electron, the entire trace is multiplied by -1 to ensure that the change on the SET current is always positive. The measured current is then subtracted by the baseline current. Each sample is digitised into either 0 or 1 by a fixed current threshold, where 1 corresponds to the time in which no electron occupies the qubit dot, while at 0 the dot is occupied by an electron. If a trace contains even a single digital 1, it will be counted as a spin-up event, independent of the number of jumps inside a single trace, so multiple current jumps inside a single readout window due to e.g. thermal excitation, will be counted the same as a single current jump. The resulting spin-up fraction is then calculated by dividing the number of traces containing a digital one by the total number of traces.

\clearpage

\section{Supplementary information}\label{SubInfo}

\subsection{Clifford and Primitive single qubit gates}\label{sup_Gates}

We follow \cite{epstein_investigating_2014} and decompose the single qubit Clifford operators by introducing the Pauli group P = $\{ I, \sigma_x, \sigma_y, \sigma_z \}$, with

\begin{align*}
I & = \begin{pmatrix}
1 &   &   &   \\
  & 1 &   &   \\
  &   & 1 &   \\
  &   &   & 1 \\
\end{pmatrix}, &
\sigma_x & = \begin{pmatrix}
1 &   &   &   \\
  & 1 &   &   \\
  &   & -1 &   \\
  &   &   & -1 \\
\end{pmatrix},\\
\sigma_y & = \begin{pmatrix}
1 &   &   &   \\
  & -1 &   &   \\
  &   & 1 &   \\
  &   &   & -1 \\
\end{pmatrix}, &
\sigma_z & = \begin{pmatrix}
1 &   &   &   \\
  & -1 &   &   \\
  &   & -1 &   \\
  &   &   & 1 \\
\end{pmatrix},
\end{align*}
where $\sigma_x$, $\sigma_y$, $\sigma_z$ correspond to $\pi$ rotations around the x, y and z axes.
The exchange group S = $\{ I,\mathcal{S},\mathcal{S}^2\}$ with

\begin{equation*}
\mathcal{S} = \begin{pmatrix}
1 &   &   &   \\
  &   &   & 1 \\
  & 1 &   &   \\
  &   & 1 &   \\
\end{pmatrix},
\mathcal{S}^2 = \begin{pmatrix}
1 &   &   &   \\
  &   & 1 &   \\
  &   &   & 1 \\
  & 1 &   &   \\
\end{pmatrix},
\end{equation*}
permutes (x,y,z) $\rightarrow$ (z,x,y) $\rightarrow$ (y,z,x), and the Hadamard group H = {$I$,$\mathcal{H}$}

\begin{equation*}
\mathcal{H} = \begin{pmatrix}
1 &   &   &   \\
  &   &   & 1 \\
  &   & -1&   \\
  & 1 &   &   \\
\end{pmatrix}
\end{equation*}
which exchanges (x,y,z) $\rightarrow$ (z,-y,x). The single qubit Clifford gates are the possible combinations of the elements in P, S and H and result in $4\,\mathrm{x}\, 3 \,\mathrm{x}\, 2 = 24$ single qubit Clifford gates. \cref{tab1} gives the complete list of all 24 single qubit Clifford gates and the physical representation using the primitive gate set \{I, $\pm$X, $\pm$Y, $\pm$X$^2$, $\pm$Y$^2$\}. With this, a single qubit Clifford gate consists on average out of 1.875 primitive gates.

\begin{table}[t]
\caption{24 Single qubit Clifford gates and primitive gate representation}\label{tab1}%
\begin{ruledtabular}
\begin{tabular}{ll}

Clifford elements  & Primitive gates\\
\midrule
$I,I,I$ & $I$ \\
$\sigma_x,I,I$ & $X^2$ \\
$\sigma_y,I,I$ & $Y^2$ \\
$\sigma_z,I,I$ & $X^2,Y^2$ \\
    &   \\
$I,I,\mathcal{S}$  & $Y,X$ \\
$\sigma_x,I,\mathcal{S}$ & $-Y,-X$ \\
$\sigma_y,I,\mathcal{S}$ & $-Y,X$ \\
$\sigma_z,I,\mathcal{S}$ & $Y,-X$ \\
    &   \\
$I,I,\mathcal{S}^2$ & $-X,-Y$ \\
$\sigma_x,I,\mathcal{S}^2$ & $X,-Y$ \\
$\sigma_y,I,\mathcal{S}^2$ & $X,Y$ \\
$\sigma_z,I,\mathcal{S}^2$ & $-X,Y$ \\
    &   \\
$I,\mathcal{H},I$ & $Y,X^2$ \\
$\sigma_x,\mathcal{H},I$ & $-Y$ \\
$\sigma_y,\mathcal{H},I$ & $-Y,X^2$ \\
$\sigma_z,\mathcal{H},I$ & $Y$ \\
    &   \\
$I,\mathcal{H},\mathcal{S}$ & $-X$ \\
$\sigma_x,\mathcal{H},\mathcal{S}$ & $X$ \\
$\sigma_y,\mathcal{H},\mathcal{S}$ & $X,Y^2$ \\
$\sigma_z,\mathcal{H},\mathcal{S}$ & $-X,Y^2$ \\
    &   \\
$I,\mathcal{H},\mathcal{S}^2$ & $X,-Y,-X$ \\
$\sigma_x,\mathcal{H},\mathcal{S}^2$ & $X,Y,X$ \\
$\sigma_y,\mathcal{H},\mathcal{S}^2$ & $X,-Y,X$ \\
$\sigma_z,\mathcal{H},\mathcal{S}^2$ & $X,Y,-X$ \\
\end{tabular}
\end{ruledtabular}
\end{table}

\newpage

\begin{figure*}[t]%
\centering
\includegraphics[width=0.60\textwidth]{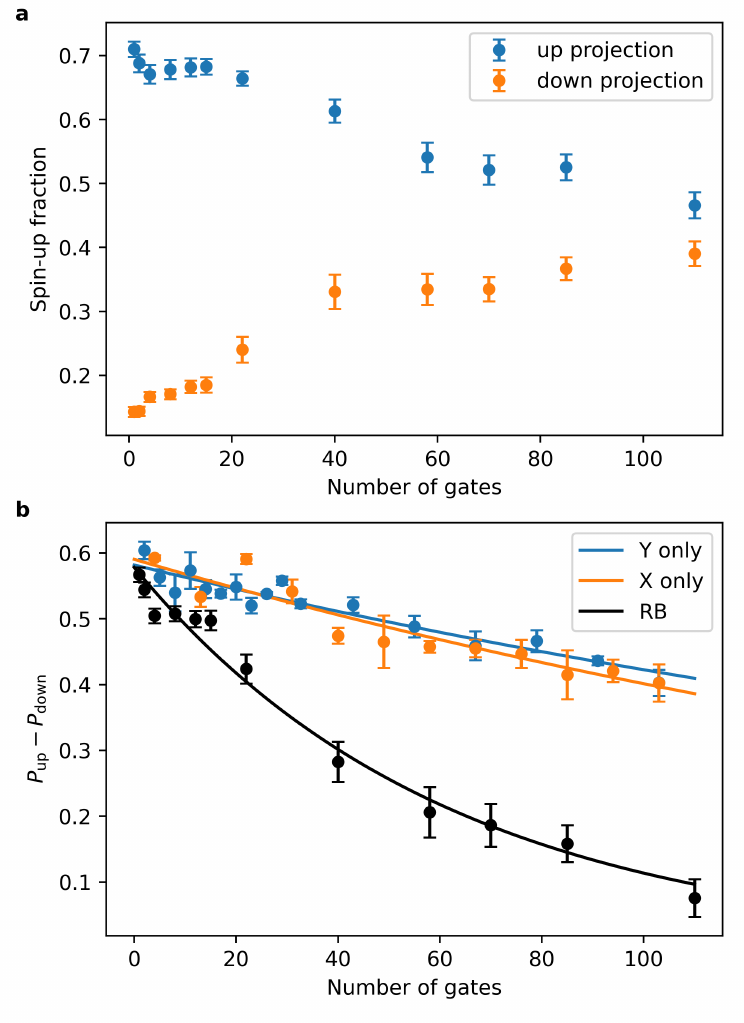}
\caption{Randomized benchmarking results for qubit A. \textbf{a} Spin-up fraction over the number of randomly applied gates with an ideal final result of spin-up (blue) and spin-down (orange). Every point corresponds to the mean spin-up fraction of 200 single shot measurements of 40 different randomized sequences. \textbf{b} Black data corresponds to the difference of the blue and orange data shown in \textbf{a} and is the final result shown in \cref{fig4}\textbf{d} in the main text resulting in a primitive gate fidelity of \SI{99.2(1)}{\percent}. Blue and orange correspond to rotations around one axis only, yielding gate fidelities of \SI{99.81(2)}{\percent} for X-only rotations, and \SI{99.82(2)}{\percent} for Y-only rotations. These fidelities are less affected by pulse calibration errors of the randomized sequences used in randomized benchmarking, and demonstrate an upper boundary set by the qubit lifetime itself.}\label{Supfig2}
\end{figure*}

\begin{figure*}[h]%
\centering
\includegraphics[width=1\textwidth]{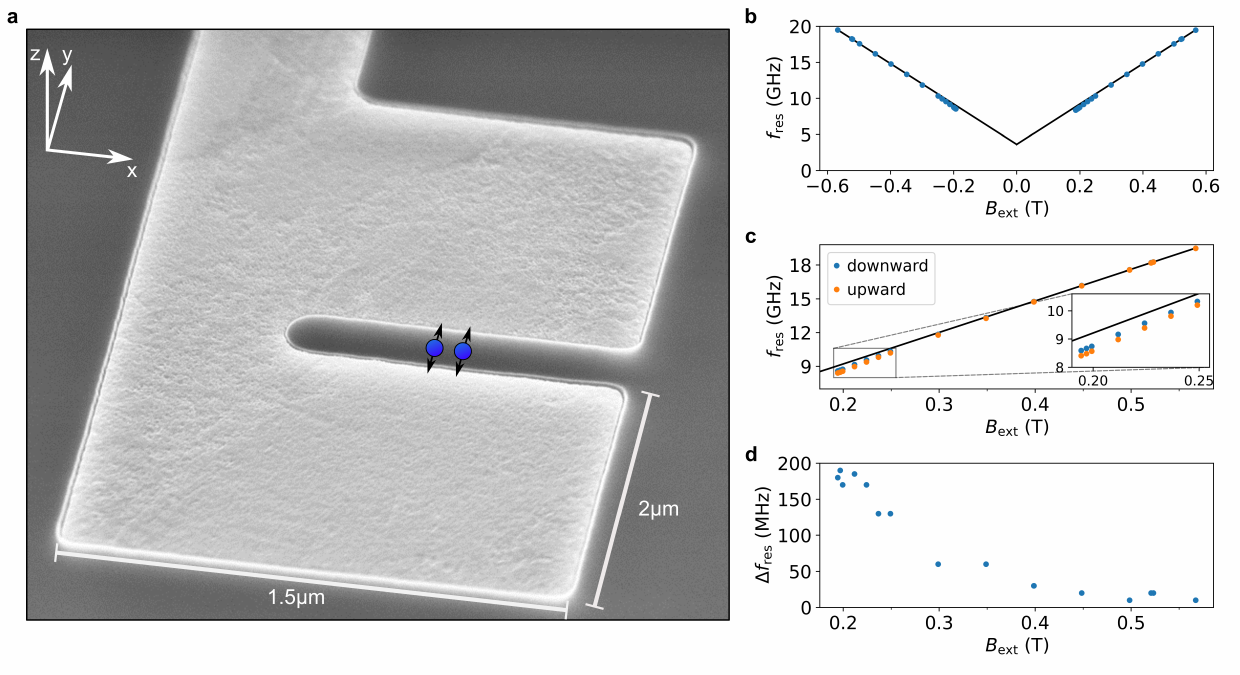}
\caption{\textbf{a} Scanning electron microscopy image of a cobalt micromagnet with a gap size of \SI{300}{nm}. The image was taken from an angled viewpoint, so the lengths in the y-direction appear smaller than in the x-direction. The Co layer has a thickness of \SI{250}{nm}, is encapsulated in SiO$_2$ and connected to a via in order to be able to ground the micromagnet. The blue dots and black arrows indicate the position and oscillation direction of the quantum dots, which are located in the quantum well approximately \SI{120}{nm} below the CoMM. \textbf{b} Resonance frequency of qubit A as a function of the applied external magnetic field in y-direction. The frequencies are extracted starting from the maximum external field and decreasing to zero. The two branches for positive and negative fields are measured independently. For high external magnetic fields, we expect the CoMM magnetisation to be fully saturated and provide a constant offset. The black lines correspond to a linear fit of all resonance frequencies between 15 and \SI{20}{GHz}. The contribution of the CoMM field in saturation is extracted from the y-offset of the linear fit and yields a field of \SI{129(2)}{\milli\tesla}. \textbf{c} Zoom on the positive magnetic field branch shown in \textbf{b}, to which qubit resonance frequencies extracted from an upward magnetic field sweep have been added. For external magnetic field values below \SI{0.3}{T}, a deviation from the linear behaviour is visible. In addition, the extracted qubit resonance frequencies deviate depending on the magnetic field sweep direction and chronology. The inset shows a zoom of the lowest measured resonance frequencies at which the deviation is the greatest. We were unable to resolve resonance frequencies below \SI{8}{GHz}, which we attribute to the decreasing magnetisation of the micromagnet and its field gradient, which are needed for EDSR drive. \textbf{d} Difference in qubit resonance frequencies between the upward and downward sweep direction of the external magnetic field. For external fields of \SI{0.2}{T}, a resonance frequency difference of $\approx$ \SI{200}{MHz} is measured, which converges to zero for externally applied magnetic fields above \SI{0.45}{T}.}\label{Supfig3}
\end{figure*}

\begin{figure*}[h]%
\centering
\includegraphics[width=0.75\textwidth]{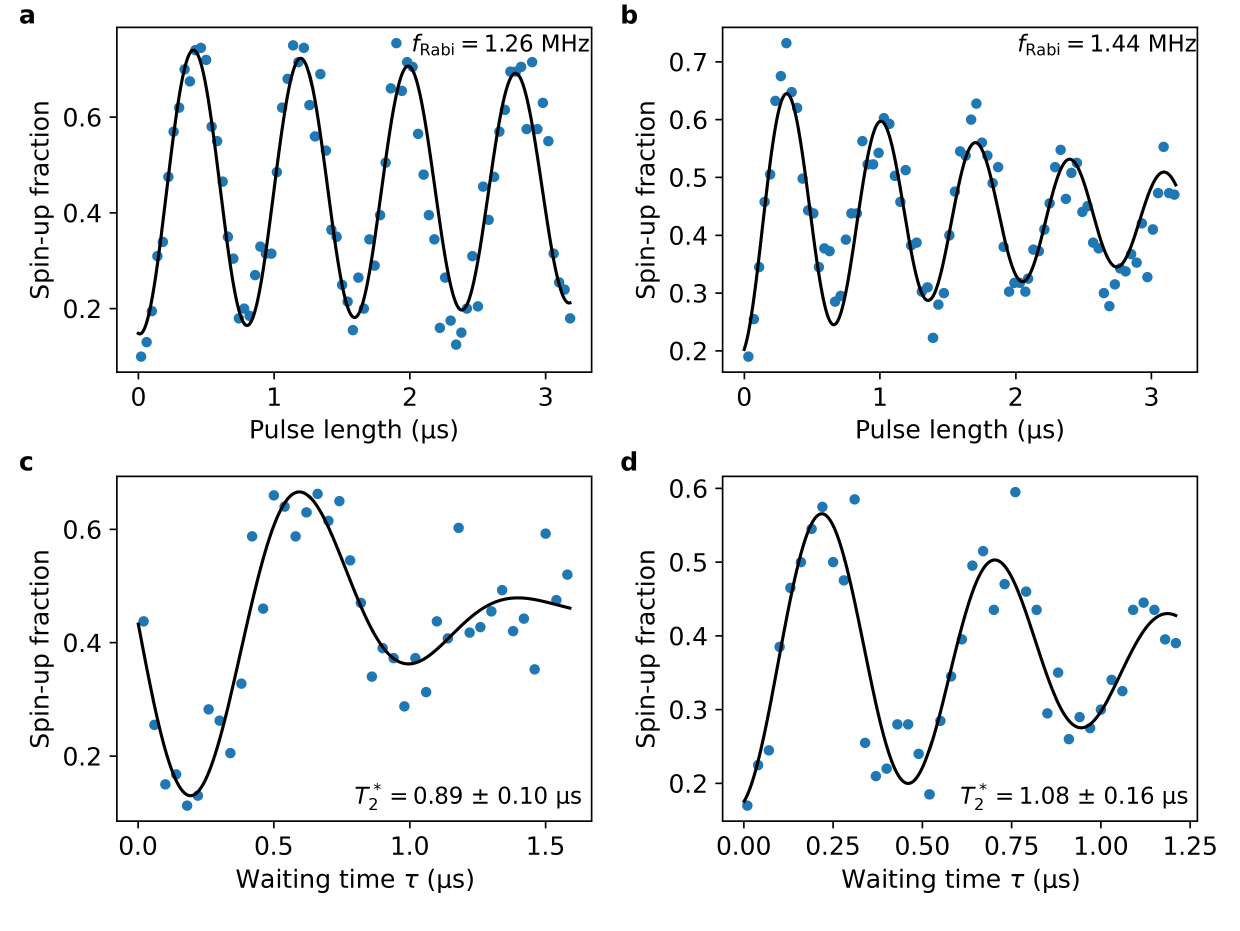}
\caption{Characterisation of qubit B and C. Both qubits samples posess a \SI{650}{\nano\metre} CoMM gap size compared to qubit A with a gap size of \SI{300}{nm}. Qubit A and C origin from the same \SI{300}{\milli\metre} wafer, while qubit B comes from a different wafer. A \SI{650}{nm} gap size investigated in order to minimize the dephasing gradient, which could limit the coherence times in enriched Si$^{28}$ devices \cite{dumoulin_stuyck_low_2021}. \textbf{a} Rabi oscillations of qubit B. The fit results in a Rabi frequency of \SI{1.26}{\mega\hertz}. \textbf{b} Rabi oscillations of qubit C. The fit results in a Rabi frequency of \SI{1.44}{\mega\hertz}.
\textbf{c} Ramsey oscillations of qubit B with a frequency detuning of \SI{1}{\mega\hertz}. A gaussian damped sine fit results in a $T_2^*$ value of \SI{0.89(0.10)}{\micro\second}. \textbf{d} Ramsey oscillations of qubit C with a frequency detuning of \SI{2}{\mega\hertz}, yielding a $T_2^*$ value of \SI{1.08(0.16)}{\micro\second}.}
\label{Supfig1}
\end{figure*}

\newpage
\clearpage

\bibliography{main}% Produces the bibliography via BibTeX.

%apsrev4-2.bst 2019-01-14 (MD) hand-edited version of apsrev4-1.bst
%Control: key (0)
%Control: author (8) initials jnrlst
%Control: editor formatted (1) identically to author
%Control: production of article title (0) allowed
%Control: page (0) single
%Control: year (1) truncated
%Control: production of eprint (0) enabled
\begin{thebibliography}{41}%
\makeatletter
\providecommand \@ifxundefined [1]{%
 \@ifx{#1\undefined}
}%
\providecommand \@ifnum [1]{%
 \ifnum #1\expandafter \@firstoftwo
 \else \expandafter \@secondoftwo
 \fi
}%
\providecommand \@ifx [1]{%
 \ifx #1\expandafter \@firstoftwo
 \else \expandafter \@secondoftwo
 \fi
}%
\providecommand \natexlab [1]{#1}%
\providecommand \enquote  [1]{``#1''}%
\providecommand \bibnamefont  [1]{#1}%
\providecommand \bibfnamefont [1]{#1}%
\providecommand \citenamefont [1]{#1}%
\providecommand \href@noop [0]{\@secondoftwo}%
\providecommand \href [0]{\begingroup \@sanitize@url \@href}%
\providecommand \@href[1]{\@@startlink{#1}\@@href}%
\providecommand \@@href[1]{\endgroup#1\@@endlink}%
\providecommand \@sanitize@url [0]{\catcode `\\12\catcode `\$12\catcode `\&12\catcode `\#12\catcode `\^12\catcode `\_12\catcode `\%12\relax}%
\providecommand \@@startlink[1]{}%
\providecommand \@@endlink[0]{}%
\providecommand \url  [0]{\begingroup\@sanitize@url \@url }%
\providecommand \@url [1]{\endgroup\@href {#1}{\urlprefix }}%
\providecommand \urlprefix  [0]{URL }%
\providecommand \Eprint [0]{\href }%
\providecommand \doibase [0]{https://doi.org/}%
\providecommand \selectlanguage [0]{\@gobble}%
\providecommand \bibinfo  [0]{\@secondoftwo}%
\providecommand \bibfield  [0]{\@secondoftwo}%
\providecommand \translation [1]{[#1]}%
\providecommand \BibitemOpen [0]{}%
\providecommand \bibitemStop [0]{}%
\providecommand \bibitemNoStop [0]{.\EOS\space}%
\providecommand \EOS [0]{\spacefactor3000\relax}%
\providecommand \BibitemShut  [1]{\csname bibitem#1\endcsname}%
\let\auto@bib@innerbib\@empty
%</preamble>
\bibitem [{\citenamefont {Loss}\ and\ \citenamefont {DiVincenzo}(1998)}]{loss_quantum_1998}%
  \BibitemOpen
  \bibfield  {author} {\bibinfo {author} {\bibfnamefont {D.}~\bibnamefont {Loss}}\ and\ \bibinfo {author} {\bibfnamefont {D.~P.}\ \bibnamefont {DiVincenzo}},\ }\bibfield  {title} {\bibinfo {title} {Quantum computation with quantum dots},\ }\href {https://doi.org/10.1103/PhysRevA.57.120} {\bibfield  {journal} {\bibinfo  {journal} {Physical Review A}\ }\textbf {\bibinfo {volume} {57}},\ \bibinfo {pages} {120} (\bibinfo {year} {1998})}\BibitemShut {NoStop}%
\bibitem [{\citenamefont {Vandersypen}\ \emph {et~al.}(2017)\citenamefont {Vandersypen}, \citenamefont {Bluhm}, \citenamefont {Clarke}, \citenamefont {Dzurak}, \citenamefont {Ishihara}, \citenamefont {Morello}, \citenamefont {Reilly}, \citenamefont {Schreiber},\ and\ \citenamefont {Veldhorst}}]{vandersypen_interfacing_2017}%
  \BibitemOpen
  \bibfield  {author} {\bibinfo {author} {\bibfnamefont {L.~M.~K.}\ \bibnamefont {Vandersypen}}, \bibinfo {author} {\bibfnamefont {H.}~\bibnamefont {Bluhm}}, \bibinfo {author} {\bibfnamefont {J.~S.}\ \bibnamefont {Clarke}}, \bibinfo {author} {\bibfnamefont {A.~S.}\ \bibnamefont {Dzurak}}, \bibinfo {author} {\bibfnamefont {R.}~\bibnamefont {Ishihara}}, \bibinfo {author} {\bibfnamefont {A.}~\bibnamefont {Morello}}, \bibinfo {author} {\bibfnamefont {D.~J.}\ \bibnamefont {Reilly}}, \bibinfo {author} {\bibfnamefont {L.~R.}\ \bibnamefont {Schreiber}},\ and\ \bibinfo {author} {\bibfnamefont {M.}~\bibnamefont {Veldhorst}},\ }\bibfield  {title} {\bibinfo {title} {Interfacing spin qubits in quantum dots and donors—hot, dense, and coherent},\ }\href {https://doi.org/10.1038/s41534-017-0038-y} {\bibfield  {journal} {\bibinfo  {journal} {npj Quantum Information}\ }\textbf {\bibinfo {volume} {3}},\ \bibinfo {pages} {34} (\bibinfo {year} {2017})}\BibitemShut {NoStop}%
\bibitem [{\citenamefont {Yoneda}\ \emph {et~al.}(2018)\citenamefont {Yoneda}, \citenamefont {Takeda}, \citenamefont {Otsuka}, \citenamefont {Nakajima}, \citenamefont {Delbecq}, \citenamefont {Allison}, \citenamefont {Honda}, \citenamefont {Kodera}, \citenamefont {Oda}, \citenamefont {Hoshi}, \citenamefont {Usami}, \citenamefont {Itoh},\ and\ \citenamefont {Tarucha}}]{yoneda_quantum-dot_2018}%
  \BibitemOpen
  \bibfield  {author} {\bibinfo {author} {\bibfnamefont {J.}~\bibnamefont {Yoneda}}, \bibinfo {author} {\bibfnamefont {K.}~\bibnamefont {Takeda}}, \bibinfo {author} {\bibfnamefont {T.}~\bibnamefont {Otsuka}}, \bibinfo {author} {\bibfnamefont {T.}~\bibnamefont {Nakajima}}, \bibinfo {author} {\bibfnamefont {M.~R.}\ \bibnamefont {Delbecq}}, \bibinfo {author} {\bibfnamefont {G.}~\bibnamefont {Allison}}, \bibinfo {author} {\bibfnamefont {T.}~\bibnamefont {Honda}}, \bibinfo {author} {\bibfnamefont {T.}~\bibnamefont {Kodera}}, \bibinfo {author} {\bibfnamefont {S.}~\bibnamefont {Oda}}, \bibinfo {author} {\bibfnamefont {Y.}~\bibnamefont {Hoshi}}, \bibinfo {author} {\bibfnamefont {N.}~\bibnamefont {Usami}}, \bibinfo {author} {\bibfnamefont {K.~M.}\ \bibnamefont {Itoh}},\ and\ \bibinfo {author} {\bibfnamefont {S.}~\bibnamefont {Tarucha}},\ }\bibfield  {title} {\bibinfo {title} {A quantum-dot spin qubit with coherence limited by charge noise and fidelity higher than 99.9\%},\ }\href
  {https://doi.org/10.1038/s41565-017-0014-x} {\bibfield  {journal} {\bibinfo  {journal} {Nature Nanotechnology}\ }\textbf {\bibinfo {volume} {13}},\ \bibinfo {pages} {102} (\bibinfo {year} {2018})}\BibitemShut {NoStop}%
\bibitem [{\citenamefont {Xue}\ \emph {et~al.}(2019)\citenamefont {Xue}, \citenamefont {Watson}, \citenamefont {Helsen}, \citenamefont {Ward}, \citenamefont {Savage}, \citenamefont {Lagally}, \citenamefont {Coppersmith}, \citenamefont {Eriksson}, \citenamefont {Wehner},\ and\ \citenamefont {Vandersypen}}]{xue_benchmarking_2019}%
  \BibitemOpen
  \bibfield  {author} {\bibinfo {author} {\bibfnamefont {X.}~\bibnamefont {Xue}}, \bibinfo {author} {\bibfnamefont {T.}~\bibnamefont {Watson}}, \bibinfo {author} {\bibfnamefont {J.}~\bibnamefont {Helsen}}, \bibinfo {author} {\bibfnamefont {D.}~\bibnamefont {Ward}}, \bibinfo {author} {\bibfnamefont {D.}~\bibnamefont {Savage}}, \bibinfo {author} {\bibfnamefont {M.}~\bibnamefont {Lagally}}, \bibinfo {author} {\bibfnamefont {S.}~\bibnamefont {Coppersmith}}, \bibinfo {author} {\bibfnamefont {M.}~\bibnamefont {Eriksson}}, \bibinfo {author} {\bibfnamefont {S.}~\bibnamefont {Wehner}},\ and\ \bibinfo {author} {\bibfnamefont {L.}~\bibnamefont {Vandersypen}},\ }\bibfield  {title} {\bibinfo {title} {Benchmarking {Gate} {Fidelities} in a {Si} / {SiGe} {Two}-{Qubit} {Device}},\ }\href {https://doi.org/10.1103/PhysRevX.9.021011} {\bibfield  {journal} {\bibinfo  {journal} {Physical Review X}\ }\textbf {\bibinfo {volume} {9}},\ \bibinfo {pages} {021011} (\bibinfo {year} {2019})}\BibitemShut {NoStop}%
\bibitem [{\citenamefont {Huang}\ \emph {et~al.}(2019)\citenamefont {Huang}, \citenamefont {Yang}, \citenamefont {Chan}, \citenamefont {Tanttu}, \citenamefont {Hensen}, \citenamefont {Leon}, \citenamefont {Fogarty}, \citenamefont {Hwang}, \citenamefont {Hudson}, \citenamefont {Itoh}, \citenamefont {Morello}, \citenamefont {Laucht},\ and\ \citenamefont {Dzurak}}]{huang_fidelity_2019}%
  \BibitemOpen
  \bibfield  {author} {\bibinfo {author} {\bibfnamefont {W.}~\bibnamefont {Huang}}, \bibinfo {author} {\bibfnamefont {C.~H.}\ \bibnamefont {Yang}}, \bibinfo {author} {\bibfnamefont {K.~W.}\ \bibnamefont {Chan}}, \bibinfo {author} {\bibfnamefont {T.}~\bibnamefont {Tanttu}}, \bibinfo {author} {\bibfnamefont {B.}~\bibnamefont {Hensen}}, \bibinfo {author} {\bibfnamefont {R.~C.~C.}\ \bibnamefont {Leon}}, \bibinfo {author} {\bibfnamefont {M.~A.}\ \bibnamefont {Fogarty}}, \bibinfo {author} {\bibfnamefont {J.~C.~C.}\ \bibnamefont {Hwang}}, \bibinfo {author} {\bibfnamefont {F.~E.}\ \bibnamefont {Hudson}}, \bibinfo {author} {\bibfnamefont {K.~M.}\ \bibnamefont {Itoh}}, \bibinfo {author} {\bibfnamefont {A.}~\bibnamefont {Morello}}, \bibinfo {author} {\bibfnamefont {A.}~\bibnamefont {Laucht}},\ and\ \bibinfo {author} {\bibfnamefont {A.~S.}\ \bibnamefont {Dzurak}},\ }\bibfield  {title} {\bibinfo {title} {Fidelity benchmarks for two-qubit gates in silicon},\ }\href {https://doi.org/10.1038/s41586-019-1197-0} {\bibfield
  {journal} {\bibinfo  {journal} {Nature}\ }\textbf {\bibinfo {volume} {569}},\ \bibinfo {pages} {532} (\bibinfo {year} {2019})}\BibitemShut {NoStop}%
\bibitem [{\citenamefont {Ono}\ \emph {et~al.}(2019)\citenamefont {Ono}, \citenamefont {Mori},\ and\ \citenamefont {Moriyama}}]{ono_high-temperature_2019}%
  \BibitemOpen
  \bibfield  {author} {\bibinfo {author} {\bibfnamefont {K.}~\bibnamefont {Ono}}, \bibinfo {author} {\bibfnamefont {T.}~\bibnamefont {Mori}},\ and\ \bibinfo {author} {\bibfnamefont {S.}~\bibnamefont {Moriyama}},\ }\bibfield  {title} {\bibinfo {title} {High-temperature operation of a silicon qubit},\ }\href {https://doi.org/10.1038/s41598-018-36476-z} {\bibfield  {journal} {\bibinfo  {journal} {Scientific Reports}\ }\textbf {\bibinfo {volume} {9}},\ \bibinfo {pages} {469} (\bibinfo {year} {2019})}\BibitemShut {NoStop}%
\bibitem [{\citenamefont {Undseth}\ \emph {et~al.}(2023)\citenamefont {Undseth}, \citenamefont {Pietx-Casas}, \citenamefont {Raymenants}, \citenamefont {Mehmandoost}, \citenamefont {Mądzik}, \citenamefont {Philips}, \citenamefont {De~Snoo}, \citenamefont {Michalak}, \citenamefont {Amitonov}, \citenamefont {Tryputen}, \citenamefont {Wuetz}, \citenamefont {Fezzi}, \citenamefont {Esposti}, \citenamefont {Sammak}, \citenamefont {Scappucci},\ and\ \citenamefont {Vandersypen}}]{undseth_hotter_2023}%
  \BibitemOpen
  \bibfield  {author} {\bibinfo {author} {\bibfnamefont {B.}~\bibnamefont {Undseth}}, \bibinfo {author} {\bibfnamefont {O.}~\bibnamefont {Pietx-Casas}}, \bibinfo {author} {\bibfnamefont {E.}~\bibnamefont {Raymenants}}, \bibinfo {author} {\bibfnamefont {M.}~\bibnamefont {Mehmandoost}}, \bibinfo {author} {\bibfnamefont {M.~T.}\ \bibnamefont {Mądzik}}, \bibinfo {author} {\bibfnamefont {S.~G.}\ \bibnamefont {Philips}}, \bibinfo {author} {\bibfnamefont {S.~L.}\ \bibnamefont {De~Snoo}}, \bibinfo {author} {\bibfnamefont {D.~J.}\ \bibnamefont {Michalak}}, \bibinfo {author} {\bibfnamefont {S.~V.}\ \bibnamefont {Amitonov}}, \bibinfo {author} {\bibfnamefont {L.}~\bibnamefont {Tryputen}}, \bibinfo {author} {\bibfnamefont {B.~P.}\ \bibnamefont {Wuetz}}, \bibinfo {author} {\bibfnamefont {V.}~\bibnamefont {Fezzi}}, \bibinfo {author} {\bibfnamefont {D.~D.}\ \bibnamefont {Esposti}}, \bibinfo {author} {\bibfnamefont {A.}~\bibnamefont {Sammak}}, \bibinfo {author} {\bibfnamefont {G.}~\bibnamefont {Scappucci}},\ and\ \bibinfo
  {author} {\bibfnamefont {L.~M.}\ \bibnamefont {Vandersypen}},\ }\bibfield  {title} {\bibinfo {title} {Hotter is {Easier}: {Unexpected} {Temperature} {Dependence} of {Spin} {Qubit} {Frequencies}},\ }\href {https://doi.org/10.1103/PhysRevX.13.041015} {\bibfield  {journal} {\bibinfo  {journal} {Physical Review X}\ }\textbf {\bibinfo {volume} {13}},\ \bibinfo {pages} {041015} (\bibinfo {year} {2023})}\BibitemShut {NoStop}%
\bibitem [{\citenamefont {Huang}\ \emph {et~al.}(2024)\citenamefont {Huang}, \citenamefont {Su}, \citenamefont {Lim}, \citenamefont {Feng}, \citenamefont {Van~Straaten}, \citenamefont {Severin}, \citenamefont {Gilbert}, \citenamefont {Dumoulin~Stuyck}, \citenamefont {Tanttu}, \citenamefont {Serrano}, \citenamefont {Cifuentes}, \citenamefont {Hansen}, \citenamefont {Seedhouse}, \citenamefont {Vahapoglu}, \citenamefont {Leon}, \citenamefont {Abrosimov}, \citenamefont {Pohl}, \citenamefont {Thewalt}, \citenamefont {Hudson}, \citenamefont {Escott}, \citenamefont {Ares}, \citenamefont {Bartlett}, \citenamefont {Morello}, \citenamefont {Saraiva}, \citenamefont {Laucht}, \citenamefont {Dzurak},\ and\ \citenamefont {Yang}}]{huang_high-fidelity_2024}%
  \BibitemOpen
  \bibfield  {author} {\bibinfo {author} {\bibfnamefont {J.~Y.}\ \bibnamefont {Huang}}, \bibinfo {author} {\bibfnamefont {R.~Y.}\ \bibnamefont {Su}}, \bibinfo {author} {\bibfnamefont {W.~H.}\ \bibnamefont {Lim}}, \bibinfo {author} {\bibfnamefont {M.}~\bibnamefont {Feng}}, \bibinfo {author} {\bibfnamefont {B.}~\bibnamefont {Van~Straaten}}, \bibinfo {author} {\bibfnamefont {B.}~\bibnamefont {Severin}}, \bibinfo {author} {\bibfnamefont {W.}~\bibnamefont {Gilbert}}, \bibinfo {author} {\bibfnamefont {N.}~\bibnamefont {Dumoulin~Stuyck}}, \bibinfo {author} {\bibfnamefont {T.}~\bibnamefont {Tanttu}}, \bibinfo {author} {\bibfnamefont {S.}~\bibnamefont {Serrano}}, \bibinfo {author} {\bibfnamefont {J.~D.}\ \bibnamefont {Cifuentes}}, \bibinfo {author} {\bibfnamefont {I.}~\bibnamefont {Hansen}}, \bibinfo {author} {\bibfnamefont {A.~E.}\ \bibnamefont {Seedhouse}}, \bibinfo {author} {\bibfnamefont {E.}~\bibnamefont {Vahapoglu}}, \bibinfo {author} {\bibfnamefont {R.~C.~C.}\ \bibnamefont {Leon}}, \bibinfo {author}
  {\bibfnamefont {N.~V.}\ \bibnamefont {Abrosimov}}, \bibinfo {author} {\bibfnamefont {H.-J.}\ \bibnamefont {Pohl}}, \bibinfo {author} {\bibfnamefont {M.~L.~W.}\ \bibnamefont {Thewalt}}, \bibinfo {author} {\bibfnamefont {F.~E.}\ \bibnamefont {Hudson}}, \bibinfo {author} {\bibfnamefont {C.~C.}\ \bibnamefont {Escott}}, \bibinfo {author} {\bibfnamefont {N.}~\bibnamefont {Ares}}, \bibinfo {author} {\bibfnamefont {S.~D.}\ \bibnamefont {Bartlett}}, \bibinfo {author} {\bibfnamefont {A.}~\bibnamefont {Morello}}, \bibinfo {author} {\bibfnamefont {A.}~\bibnamefont {Saraiva}}, \bibinfo {author} {\bibfnamefont {A.}~\bibnamefont {Laucht}}, \bibinfo {author} {\bibfnamefont {A.~S.}\ \bibnamefont {Dzurak}},\ and\ \bibinfo {author} {\bibfnamefont {C.~H.}\ \bibnamefont {Yang}},\ }\bibfield  {title} {\bibinfo {title} {High-fidelity spin qubit operation and algorithmic initialization above 1 {K}},\ }\href {https://doi.org/10.1038/s41586-024-07160-2} {\bibfield  {journal} {\bibinfo  {journal} {Nature}\ }\textbf {\bibinfo {volume}
  {627}},\ \bibinfo {pages} {772} (\bibinfo {year} {2024})}\BibitemShut {NoStop}%
\bibitem [{\citenamefont {Seidler}\ \emph {et~al.}(2022)\citenamefont {Seidler}, \citenamefont {Struck}, \citenamefont {Xue}, \citenamefont {Focke}, \citenamefont {Trellenkamp}, \citenamefont {Bluhm},\ and\ \citenamefont {Schreiber}}]{seidler_conveyor-mode_2022}%
  \BibitemOpen
  \bibfield  {author} {\bibinfo {author} {\bibfnamefont {I.}~\bibnamefont {Seidler}}, \bibinfo {author} {\bibfnamefont {T.}~\bibnamefont {Struck}}, \bibinfo {author} {\bibfnamefont {R.}~\bibnamefont {Xue}}, \bibinfo {author} {\bibfnamefont {N.}~\bibnamefont {Focke}}, \bibinfo {author} {\bibfnamefont {S.}~\bibnamefont {Trellenkamp}}, \bibinfo {author} {\bibfnamefont {H.}~\bibnamefont {Bluhm}},\ and\ \bibinfo {author} {\bibfnamefont {L.~R.}\ \bibnamefont {Schreiber}},\ }\bibfield  {title} {\bibinfo {title} {Conveyor-mode single-electron shuttling in {Si}/{SiGe} for a scalable quantum computing architecture},\ }\href {https://doi.org/10.1038/s41534-022-00615-2} {\bibfield  {journal} {\bibinfo  {journal} {npj Quantum Information}\ }\textbf {\bibinfo {volume} {8}},\ \bibinfo {pages} {100} (\bibinfo {year} {2022})}\BibitemShut {NoStop}%
\bibitem [{\citenamefont {Dijkema}\ \emph {et~al.}(2023)\citenamefont {Dijkema}, \citenamefont {Xue}, \citenamefont {Harvey-Collard}, \citenamefont {Rimbach-Russ}, \citenamefont {de~Snoo}, \citenamefont {Zheng}, \citenamefont {Sammak}, \citenamefont {Scappucci},\ and\ \citenamefont {Vandersypen}}]{dijkema_two-qubit_2023}%
  \BibitemOpen
  \bibfield  {author} {\bibinfo {author} {\bibfnamefont {J.}~\bibnamefont {Dijkema}}, \bibinfo {author} {\bibfnamefont {X.}~\bibnamefont {Xue}}, \bibinfo {author} {\bibfnamefont {P.}~\bibnamefont {Harvey-Collard}}, \bibinfo {author} {\bibfnamefont {M.}~\bibnamefont {Rimbach-Russ}}, \bibinfo {author} {\bibfnamefont {S.~L.}\ \bibnamefont {de~Snoo}}, \bibinfo {author} {\bibfnamefont {G.}~\bibnamefont {Zheng}}, \bibinfo {author} {\bibfnamefont {A.}~\bibnamefont {Sammak}}, \bibinfo {author} {\bibfnamefont {G.}~\bibnamefont {Scappucci}},\ and\ \bibinfo {author} {\bibfnamefont {L.~M.~K.}\ \bibnamefont {Vandersypen}},\ }\href {http://arxiv.org/abs/2310.16805} {\bibinfo {title} {Two-qubit logic between distant spins in silicon}} (\bibinfo {year} {2023}),\ \bibinfo {note} {arXiv:2310.16805 [cond-mat, physics:quant-ph]}\BibitemShut {NoStop}%
\bibitem [{\citenamefont {De~Smet}\ \emph {et~al.}(2024)\citenamefont {De~Smet}, \citenamefont {Matsumoto}, \citenamefont {Zwerver}, \citenamefont {Tryputen}, \citenamefont {de~Snoo}, \citenamefont {Amitonov}, \citenamefont {Sammak}, \citenamefont {Samkharadze}, \citenamefont {G{\"u}l}, \citenamefont {Wasserman}, \citenamefont {Rimbach-Russ}, \citenamefont {Scappucci},\ and\ \citenamefont {Vandersypen}}]{de_smet_high-fidelity_2024}%
  \BibitemOpen
  \bibfield  {author} {\bibinfo {author} {\bibfnamefont {M.}~\bibnamefont {De~Smet}}, \bibinfo {author} {\bibfnamefont {Y.}~\bibnamefont {Matsumoto}}, \bibinfo {author} {\bibfnamefont {A.-M.~J.}\ \bibnamefont {Zwerver}}, \bibinfo {author} {\bibfnamefont {L.}~\bibnamefont {Tryputen}}, \bibinfo {author} {\bibfnamefont {S.~L.}\ \bibnamefont {de~Snoo}}, \bibinfo {author} {\bibfnamefont {S.~V.}\ \bibnamefont {Amitonov}}, \bibinfo {author} {\bibfnamefont {A.}~\bibnamefont {Sammak}}, \bibinfo {author} {\bibfnamefont {N.}~\bibnamefont {Samkharadze}}, \bibinfo {author} {\bibfnamefont {{\"O}.}~\bibnamefont {G{\"u}l}}, \bibinfo {author} {\bibfnamefont {R.~N.~M.}\ \bibnamefont {Wasserman}}, \bibinfo {author} {\bibfnamefont {M.}~\bibnamefont {Rimbach-Russ}}, \bibinfo {author} {\bibfnamefont {G.}~\bibnamefont {Scappucci}},\ and\ \bibinfo {author} {\bibfnamefont {L.~M.~K.}\ \bibnamefont {Vandersypen}},\ }\href {http://arxiv.org/abs/2406.07267} {\bibinfo {title} {High-fidelity single-spin shuttling in silicon}} (\bibinfo
  {year} {2024}),\ \bibinfo {note} {arXiv:2406.07267 [cond-mat, physics:quant-ph]}\BibitemShut {NoStop}%
\bibitem [{\citenamefont {Philips}\ \emph {et~al.}(2022)\citenamefont {Philips}, \citenamefont {Mądzik}, \citenamefont {Amitonov}, \citenamefont {De~Snoo}, \citenamefont {Russ}, \citenamefont {Kalhor}, \citenamefont {Volk}, \citenamefont {Lawrie}, \citenamefont {Brousse}, \citenamefont {Tryputen}, \citenamefont {Wuetz}, \citenamefont {Sammak}, \citenamefont {Veldhorst}, \citenamefont {Scappucci},\ and\ \citenamefont {Vandersypen}}]{philips_universal_2022}%
  \BibitemOpen
  \bibfield  {author} {\bibinfo {author} {\bibfnamefont {S.~G.~J.}\ \bibnamefont {Philips}}, \bibinfo {author} {\bibfnamefont {M.~T.}\ \bibnamefont {Mądzik}}, \bibinfo {author} {\bibfnamefont {S.~V.}\ \bibnamefont {Amitonov}}, \bibinfo {author} {\bibfnamefont {S.~L.}\ \bibnamefont {De~Snoo}}, \bibinfo {author} {\bibfnamefont {M.}~\bibnamefont {Russ}}, \bibinfo {author} {\bibfnamefont {N.}~\bibnamefont {Kalhor}}, \bibinfo {author} {\bibfnamefont {C.}~\bibnamefont {Volk}}, \bibinfo {author} {\bibfnamefont {W.~I.~L.}\ \bibnamefont {Lawrie}}, \bibinfo {author} {\bibfnamefont {D.}~\bibnamefont {Brousse}}, \bibinfo {author} {\bibfnamefont {L.}~\bibnamefont {Tryputen}}, \bibinfo {author} {\bibfnamefont {B.~P.}\ \bibnamefont {Wuetz}}, \bibinfo {author} {\bibfnamefont {A.}~\bibnamefont {Sammak}}, \bibinfo {author} {\bibfnamefont {M.}~\bibnamefont {Veldhorst}}, \bibinfo {author} {\bibfnamefont {G.}~\bibnamefont {Scappucci}},\ and\ \bibinfo {author} {\bibfnamefont {L.~M.~K.}\ \bibnamefont {Vandersypen}},\ }\bibfield
  {title} {\bibinfo {title} {Universal control of a six-qubit quantum processor in silicon},\ }\href {https://doi.org/10.1038/s41586-022-05117-x} {\bibfield  {journal} {\bibinfo  {journal} {Nature}\ }\textbf {\bibinfo {volume} {609}},\ \bibinfo {pages} {919} (\bibinfo {year} {2022})}\BibitemShut {NoStop}%
\bibitem [{\citenamefont {Shehata}\ \emph {et~al.}(2023)\citenamefont {Shehata}, \citenamefont {Simion}, \citenamefont {Li}, \citenamefont {Mohiyaddin}, \citenamefont {Wan}, \citenamefont {Mongillo}, \citenamefont {Govoreanu}, \citenamefont {Radu}, \citenamefont {De~Greve},\ and\ \citenamefont {Van~Dorpe}}]{shehata_modeling_2023}%
  \BibitemOpen
  \bibfield  {author} {\bibinfo {author} {\bibfnamefont {M.~M. E.~K.}\ \bibnamefont {Shehata}}, \bibinfo {author} {\bibfnamefont {G.}~\bibnamefont {Simion}}, \bibinfo {author} {\bibfnamefont {R.}~\bibnamefont {Li}}, \bibinfo {author} {\bibfnamefont {F.~A.}\ \bibnamefont {Mohiyaddin}}, \bibinfo {author} {\bibfnamefont {D.}~\bibnamefont {Wan}}, \bibinfo {author} {\bibfnamefont {M.}~\bibnamefont {Mongillo}}, \bibinfo {author} {\bibfnamefont {B.}~\bibnamefont {Govoreanu}}, \bibinfo {author} {\bibfnamefont {I.}~\bibnamefont {Radu}}, \bibinfo {author} {\bibfnamefont {K.}~\bibnamefont {De~Greve}},\ and\ \bibinfo {author} {\bibfnamefont {P.}~\bibnamefont {Van~Dorpe}},\ }\bibfield  {title} {\bibinfo {title} {Modeling semiconductor spin qubits and their charge noise environment for quantum gate fidelity estimation},\ }\href {https://doi.org/10.1103/PhysRevB.108.045305} {\bibfield  {journal} {\bibinfo  {journal} {Physical Review B}\ }\textbf {\bibinfo {volume} {108}},\ \bibinfo {pages} {045305} (\bibinfo {year}
  {2023})}\BibitemShut {NoStop}%
\bibitem [{\citenamefont {Fleetwood}\ \emph {et~al.}(1993)\citenamefont {Fleetwood}, \citenamefont {Winokur}, \citenamefont {Reber}, \citenamefont {Meisenheimer}, \citenamefont {Schwank}, \citenamefont {Shaneyfelt},\ and\ \citenamefont {Riewe}}]{fleetwood_effects_1993}%
  \BibitemOpen
  \bibfield  {author} {\bibinfo {author} {\bibfnamefont {D.~M.}\ \bibnamefont {Fleetwood}}, \bibinfo {author} {\bibfnamefont {P.~S.}\ \bibnamefont {Winokur}}, \bibinfo {author} {\bibfnamefont {R.~A.}\ \bibnamefont {Reber}}, \bibinfo {author} {\bibfnamefont {T.~L.}\ \bibnamefont {Meisenheimer}}, \bibinfo {author} {\bibfnamefont {J.~R.}\ \bibnamefont {Schwank}}, \bibinfo {author} {\bibfnamefont {M.~R.}\ \bibnamefont {Shaneyfelt}},\ and\ \bibinfo {author} {\bibfnamefont {L.~C.}\ \bibnamefont {Riewe}},\ }\bibfield  {title} {\bibinfo {title} {Effects of oxide traps, interface traps, and ‘‘border traps’’ on metal-oxide-semiconductor devices},\ }\href {https://doi.org/10.1063/1.353777} {\bibfield  {journal} {\bibinfo  {journal} {Journal of Applied Physics}\ }\textbf {\bibinfo {volume} {73}},\ \bibinfo {pages} {5058} (\bibinfo {year} {1993})}\BibitemShut {NoStop}%
\bibitem [{\citenamefont {Dutta}\ and\ \citenamefont {Horn}(1981)}]{dutta_low-frequency_1981}%
  \BibitemOpen
  \bibfield  {author} {\bibinfo {author} {\bibfnamefont {P.}~\bibnamefont {Dutta}}\ and\ \bibinfo {author} {\bibfnamefont {P.~M.}\ \bibnamefont {Horn}},\ }\bibfield  {title} {\bibinfo {title} {Low-frequency fluctuations in solids: 1 f noise},\ }\href {https://doi.org/10.1103/RevModPhys.53.497} {\bibfield  {journal} {\bibinfo  {journal} {Reviews of Modern Physics}\ }\textbf {\bibinfo {volume} {53}},\ \bibinfo {pages} {497} (\bibinfo {year} {1981})}\BibitemShut {NoStop}%
\bibitem [{\citenamefont {Rocki}\ \emph {et~al.}(2020)\citenamefont {Rocki}, \citenamefont {Van~Essendelft}, \citenamefont {Sharapov}, \citenamefont {Schreiber}, \citenamefont {Morrison}, \citenamefont {Kibardin}, \citenamefont {Portnoy}, \citenamefont {Dietiker}, \citenamefont {Syamlal},\ and\ \citenamefont {James}}]{rocki_fast_2020}%
  \BibitemOpen
  \bibfield  {author} {\bibinfo {author} {\bibfnamefont {K.}~\bibnamefont {Rocki}}, \bibinfo {author} {\bibfnamefont {D.}~\bibnamefont {Van~Essendelft}}, \bibinfo {author} {\bibfnamefont {I.}~\bibnamefont {Sharapov}}, \bibinfo {author} {\bibfnamefont {R.}~\bibnamefont {Schreiber}}, \bibinfo {author} {\bibfnamefont {M.}~\bibnamefont {Morrison}}, \bibinfo {author} {\bibfnamefont {V.}~\bibnamefont {Kibardin}}, \bibinfo {author} {\bibfnamefont {A.}~\bibnamefont {Portnoy}}, \bibinfo {author} {\bibfnamefont {J.~F.}\ \bibnamefont {Dietiker}}, \bibinfo {author} {\bibfnamefont {M.}~\bibnamefont {Syamlal}},\ and\ \bibinfo {author} {\bibfnamefont {M.}~\bibnamefont {James}},\ }\href {http://arxiv.org/abs/2010.03660} {\bibinfo {title} {Fast {Stencil}-{Code} {Computation} on a {Wafer}-{Scale} {Processor}}} (\bibinfo {year} {2020}),\ \bibinfo {note} {arXiv:2010.03660 [cs]}\BibitemShut {NoStop}%
\bibitem [{\citenamefont {Maurand}\ \emph {et~al.}(2016)\citenamefont {Maurand}, \citenamefont {Jehl}, \citenamefont {Kotekar-Patil}, \citenamefont {Corna}, \citenamefont {Bohuslavskyi}, \citenamefont {Laviéville}, \citenamefont {Hutin}, \citenamefont {Barraud}, \citenamefont {Vinet}, \citenamefont {Sanquer},\ and\ \citenamefont {De~Franceschi}}]{maurand_cmos_2016}%
  \BibitemOpen
  \bibfield  {author} {\bibinfo {author} {\bibfnamefont {R.}~\bibnamefont {Maurand}}, \bibinfo {author} {\bibfnamefont {X.}~\bibnamefont {Jehl}}, \bibinfo {author} {\bibfnamefont {D.}~\bibnamefont {Kotekar-Patil}}, \bibinfo {author} {\bibfnamefont {A.}~\bibnamefont {Corna}}, \bibinfo {author} {\bibfnamefont {H.}~\bibnamefont {Bohuslavskyi}}, \bibinfo {author} {\bibfnamefont {R.}~\bibnamefont {Laviéville}}, \bibinfo {author} {\bibfnamefont {L.}~\bibnamefont {Hutin}}, \bibinfo {author} {\bibfnamefont {S.}~\bibnamefont {Barraud}}, \bibinfo {author} {\bibfnamefont {M.}~\bibnamefont {Vinet}}, \bibinfo {author} {\bibfnamefont {M.}~\bibnamefont {Sanquer}},\ and\ \bibinfo {author} {\bibfnamefont {S.}~\bibnamefont {De~Franceschi}},\ }\bibfield  {title} {\bibinfo {title} {A {CMOS} silicon spin qubit},\ }\href {https://doi.org/10.1038/ncomms13575} {\bibfield  {journal} {\bibinfo  {journal} {Nature Communications}\ }\textbf {\bibinfo {volume} {7}},\ \bibinfo {pages} {13575} (\bibinfo {year} {2016})}\BibitemShut
  {NoStop}%
\bibitem [{\citenamefont {Zwerver}\ \emph {et~al.}(2022)\citenamefont {Zwerver}, \citenamefont {Krähenmann}, \citenamefont {Watson}, \citenamefont {Lampert}, \citenamefont {George}, \citenamefont {Pillarisetty}, \citenamefont {Bojarski}, \citenamefont {Amin}, \citenamefont {Amitonov}, \citenamefont {Boter}, \citenamefont {Caudillo}, \citenamefont {Correas-Serrano}, \citenamefont {Dehollain}, \citenamefont {Droulers}, \citenamefont {Henry}, \citenamefont {Kotlyar}, \citenamefont {Lodari}, \citenamefont {Lüthi}, \citenamefont {Michalak}, \citenamefont {Mueller}, \citenamefont {Neyens}, \citenamefont {Roberts}, \citenamefont {Samkharadze}, \citenamefont {Zheng}, \citenamefont {Zietz}, \citenamefont {Scappucci}, \citenamefont {Veldhorst}, \citenamefont {Vandersypen},\ and\ \citenamefont {Clarke}}]{zwerver_qubits_2022}%
  \BibitemOpen
  \bibfield  {author} {\bibinfo {author} {\bibfnamefont {A.~M.~J.}\ \bibnamefont {Zwerver}}, \bibinfo {author} {\bibfnamefont {T.}~\bibnamefont {Krähenmann}}, \bibinfo {author} {\bibfnamefont {T.~F.}\ \bibnamefont {Watson}}, \bibinfo {author} {\bibfnamefont {L.}~\bibnamefont {Lampert}}, \bibinfo {author} {\bibfnamefont {H.~C.}\ \bibnamefont {George}}, \bibinfo {author} {\bibfnamefont {R.}~\bibnamefont {Pillarisetty}}, \bibinfo {author} {\bibfnamefont {S.~A.}\ \bibnamefont {Bojarski}}, \bibinfo {author} {\bibfnamefont {P.}~\bibnamefont {Amin}}, \bibinfo {author} {\bibfnamefont {S.~V.}\ \bibnamefont {Amitonov}}, \bibinfo {author} {\bibfnamefont {J.~M.}\ \bibnamefont {Boter}}, \bibinfo {author} {\bibfnamefont {R.}~\bibnamefont {Caudillo}}, \bibinfo {author} {\bibfnamefont {D.}~\bibnamefont {Correas-Serrano}}, \bibinfo {author} {\bibfnamefont {J.~P.}\ \bibnamefont {Dehollain}}, \bibinfo {author} {\bibfnamefont {G.}~\bibnamefont {Droulers}}, \bibinfo {author} {\bibfnamefont {E.~M.}\ \bibnamefont {Henry}},
  \bibinfo {author} {\bibfnamefont {R.}~\bibnamefont {Kotlyar}}, \bibinfo {author} {\bibfnamefont {M.}~\bibnamefont {Lodari}}, \bibinfo {author} {\bibfnamefont {F.}~\bibnamefont {Lüthi}}, \bibinfo {author} {\bibfnamefont {D.~J.}\ \bibnamefont {Michalak}}, \bibinfo {author} {\bibfnamefont {B.~K.}\ \bibnamefont {Mueller}}, \bibinfo {author} {\bibfnamefont {S.}~\bibnamefont {Neyens}}, \bibinfo {author} {\bibfnamefont {J.}~\bibnamefont {Roberts}}, \bibinfo {author} {\bibfnamefont {N.}~\bibnamefont {Samkharadze}}, \bibinfo {author} {\bibfnamefont {G.}~\bibnamefont {Zheng}}, \bibinfo {author} {\bibfnamefont {O.~K.}\ \bibnamefont {Zietz}}, \bibinfo {author} {\bibfnamefont {G.}~\bibnamefont {Scappucci}}, \bibinfo {author} {\bibfnamefont {M.}~\bibnamefont {Veldhorst}}, \bibinfo {author} {\bibfnamefont {L.~M.~K.}\ \bibnamefont {Vandersypen}},\ and\ \bibinfo {author} {\bibfnamefont {J.~S.}\ \bibnamefont {Clarke}},\ }\bibfield  {title} {\bibinfo {title} {Qubits made by advanced semiconductor manufacturing},\ }\href
  {https://doi.org/10.1038/s41928-022-00727-9} {\bibfield  {journal} {\bibinfo  {journal} {Nature Electronics}\ }\textbf {\bibinfo {volume} {5}},\ \bibinfo {pages} {184} (\bibinfo {year} {2022})}\BibitemShut {NoStop}%
\bibitem [{\citenamefont {Klemt}\ \emph {et~al.}(2023)\citenamefont {Klemt}, \citenamefont {Elhomsy}, \citenamefont {Nurizzo}, \citenamefont {Hamonic}, \citenamefont {Martinez}, \citenamefont {Cardoso~Paz}, \citenamefont {Spence}, \citenamefont {Dartiailh}, \citenamefont {Jadot}, \citenamefont {Chanrion}, \citenamefont {Thiney}, \citenamefont {Lethiecq}, \citenamefont {Bertrand}, \citenamefont {Niebojewski}, \citenamefont {Bäuerle}, \citenamefont {Vinet}, \citenamefont {Niquet}, \citenamefont {Meunier},\ and\ \citenamefont {Urdampilleta}}]{klemt_electrical_2023}%
  \BibitemOpen
  \bibfield  {author} {\bibinfo {author} {\bibfnamefont {B.}~\bibnamefont {Klemt}}, \bibinfo {author} {\bibfnamefont {V.}~\bibnamefont {Elhomsy}}, \bibinfo {author} {\bibfnamefont {M.}~\bibnamefont {Nurizzo}}, \bibinfo {author} {\bibfnamefont {P.}~\bibnamefont {Hamonic}}, \bibinfo {author} {\bibfnamefont {B.}~\bibnamefont {Martinez}}, \bibinfo {author} {\bibfnamefont {B.}~\bibnamefont {Cardoso~Paz}}, \bibinfo {author} {\bibfnamefont {C.}~\bibnamefont {Spence}}, \bibinfo {author} {\bibfnamefont {M.~C.}\ \bibnamefont {Dartiailh}}, \bibinfo {author} {\bibfnamefont {B.}~\bibnamefont {Jadot}}, \bibinfo {author} {\bibfnamefont {E.}~\bibnamefont {Chanrion}}, \bibinfo {author} {\bibfnamefont {V.}~\bibnamefont {Thiney}}, \bibinfo {author} {\bibfnamefont {R.}~\bibnamefont {Lethiecq}}, \bibinfo {author} {\bibfnamefont {B.}~\bibnamefont {Bertrand}}, \bibinfo {author} {\bibfnamefont {H.}~\bibnamefont {Niebojewski}}, \bibinfo {author} {\bibfnamefont {C.}~\bibnamefont {Bäuerle}}, \bibinfo {author} {\bibfnamefont
  {M.}~\bibnamefont {Vinet}}, \bibinfo {author} {\bibfnamefont {Y.-M.}\ \bibnamefont {Niquet}}, \bibinfo {author} {\bibfnamefont {T.}~\bibnamefont {Meunier}},\ and\ \bibinfo {author} {\bibfnamefont {M.}~\bibnamefont {Urdampilleta}},\ }\bibfield  {title} {\bibinfo {title} {Electrical manipulation of a single electron spin in {CMOS} using a micromagnet and spin-valley coupling},\ }\href {https://doi.org/10.1038/s41534-023-00776-8} {\bibfield  {journal} {\bibinfo  {journal} {npj Quantum Information}\ }\textbf {\bibinfo {volume} {9}},\ \bibinfo {pages} {107} (\bibinfo {year} {2023})}\BibitemShut {NoStop}%
\bibitem [{\citenamefont {Neyens}\ \emph {et~al.}(2024)\citenamefont {Neyens}, \citenamefont {Zietz}, \citenamefont {Watson}, \citenamefont {Luthi}, \citenamefont {Nethwewala}, \citenamefont {George}, \citenamefont {Henry}, \citenamefont {Islam}, \citenamefont {Wagner}, \citenamefont {Borjans}, \citenamefont {Connors}, \citenamefont {Corrigan}, \citenamefont {Curry}, \citenamefont {Keith}, \citenamefont {Kotlyar}, \citenamefont {Lampert}, \citenamefont {Mądzik}, \citenamefont {Millard}, \citenamefont {Mohiyaddin}, \citenamefont {Pellerano}, \citenamefont {Pillarisetty}, \citenamefont {Ramsey}, \citenamefont {Savytskyy}, \citenamefont {Schaal}, \citenamefont {Zheng}, \citenamefont {Ziegler}, \citenamefont {Bishop}, \citenamefont {Bojarski}, \citenamefont {Roberts},\ and\ \citenamefont {Clarke}}]{neyens_probing_2024}%
  \BibitemOpen
  \bibfield  {author} {\bibinfo {author} {\bibfnamefont {S.}~\bibnamefont {Neyens}}, \bibinfo {author} {\bibfnamefont {O.~K.}\ \bibnamefont {Zietz}}, \bibinfo {author} {\bibfnamefont {T.~F.}\ \bibnamefont {Watson}}, \bibinfo {author} {\bibfnamefont {F.}~\bibnamefont {Luthi}}, \bibinfo {author} {\bibfnamefont {A.}~\bibnamefont {Nethwewala}}, \bibinfo {author} {\bibfnamefont {H.~C.}\ \bibnamefont {George}}, \bibinfo {author} {\bibfnamefont {E.}~\bibnamefont {Henry}}, \bibinfo {author} {\bibfnamefont {M.}~\bibnamefont {Islam}}, \bibinfo {author} {\bibfnamefont {A.~J.}\ \bibnamefont {Wagner}}, \bibinfo {author} {\bibfnamefont {F.}~\bibnamefont {Borjans}}, \bibinfo {author} {\bibfnamefont {E.~J.}\ \bibnamefont {Connors}}, \bibinfo {author} {\bibfnamefont {J.}~\bibnamefont {Corrigan}}, \bibinfo {author} {\bibfnamefont {M.~J.}\ \bibnamefont {Curry}}, \bibinfo {author} {\bibfnamefont {D.}~\bibnamefont {Keith}}, \bibinfo {author} {\bibfnamefont {R.}~\bibnamefont {Kotlyar}}, \bibinfo {author} {\bibfnamefont {L.~F.}\
  \bibnamefont {Lampert}}, \bibinfo {author} {\bibfnamefont {M.~T.}\ \bibnamefont {Mądzik}}, \bibinfo {author} {\bibfnamefont {K.}~\bibnamefont {Millard}}, \bibinfo {author} {\bibfnamefont {F.~A.}\ \bibnamefont {Mohiyaddin}}, \bibinfo {author} {\bibfnamefont {S.}~\bibnamefont {Pellerano}}, \bibinfo {author} {\bibfnamefont {R.}~\bibnamefont {Pillarisetty}}, \bibinfo {author} {\bibfnamefont {M.}~\bibnamefont {Ramsey}}, \bibinfo {author} {\bibfnamefont {R.}~\bibnamefont {Savytskyy}}, \bibinfo {author} {\bibfnamefont {S.}~\bibnamefont {Schaal}}, \bibinfo {author} {\bibfnamefont {G.}~\bibnamefont {Zheng}}, \bibinfo {author} {\bibfnamefont {J.}~\bibnamefont {Ziegler}}, \bibinfo {author} {\bibfnamefont {N.~C.}\ \bibnamefont {Bishop}}, \bibinfo {author} {\bibfnamefont {S.}~\bibnamefont {Bojarski}}, \bibinfo {author} {\bibfnamefont {J.}~\bibnamefont {Roberts}},\ and\ \bibinfo {author} {\bibfnamefont {J.~S.}\ \bibnamefont {Clarke}},\ }\bibfield  {title} {\bibinfo {title} {Probing single electrons across 300-mm spin
  qubit wafers},\ }\href {https://doi.org/10.1038/s41586-024-07275-6} {\bibfield  {journal} {\bibinfo  {journal} {Nature}\ }\textbf {\bibinfo {volume} {629}},\ \bibinfo {pages} {80} (\bibinfo {year} {2024})}\BibitemShut {NoStop}%
\bibitem [{\citenamefont {Pioro-Ladrière}\ \emph {et~al.}(2008)\citenamefont {Pioro-Ladrière}, \citenamefont {Obata}, \citenamefont {Tokura}, \citenamefont {Shin}, \citenamefont {Kubo}, \citenamefont {Yoshida}, \citenamefont {Taniyama},\ and\ \citenamefont {Tarucha}}]{pioro-ladriere_electrically_2008}%
  \BibitemOpen
  \bibfield  {author} {\bibinfo {author} {\bibfnamefont {M.}~\bibnamefont {Pioro-Ladrière}}, \bibinfo {author} {\bibfnamefont {T.}~\bibnamefont {Obata}}, \bibinfo {author} {\bibfnamefont {Y.}~\bibnamefont {Tokura}}, \bibinfo {author} {\bibfnamefont {Y.-S.}\ \bibnamefont {Shin}}, \bibinfo {author} {\bibfnamefont {T.}~\bibnamefont {Kubo}}, \bibinfo {author} {\bibfnamefont {K.}~\bibnamefont {Yoshida}}, \bibinfo {author} {\bibfnamefont {T.}~\bibnamefont {Taniyama}},\ and\ \bibinfo {author} {\bibfnamefont {S.}~\bibnamefont {Tarucha}},\ }\bibfield  {title} {\bibinfo {title} {Electrically driven single-electron spin resonance in a slanting {Zeeman} field},\ }\href {https://doi.org/10.1038/nphys1053} {\bibfield  {journal} {\bibinfo  {journal} {Nature Physics}\ }\textbf {\bibinfo {volume} {4}},\ \bibinfo {pages} {776} (\bibinfo {year} {2008})}\BibitemShut {NoStop}%
\bibitem [{\citenamefont {Elsayed}\ \emph {et~al.}(2024)\citenamefont {Elsayed}, \citenamefont {Shehata}, \citenamefont {Godfrin}, \citenamefont {Kubicek}, \citenamefont {Massar}, \citenamefont {Canvel}, \citenamefont {Jussot}, \citenamefont {Simion}, \citenamefont {Mongillo}, \citenamefont {Wan}, \citenamefont {Govoreanu}, \citenamefont {Radu}, \citenamefont {Li}, \citenamefont {Van~Dorpe},\ and\ \citenamefont {De~Greve}}]{elsayed_low_2024}%
  \BibitemOpen
  \bibfield  {author} {\bibinfo {author} {\bibfnamefont {A.}~\bibnamefont {Elsayed}}, \bibinfo {author} {\bibfnamefont {M.~M.~K.}\ \bibnamefont {Shehata}}, \bibinfo {author} {\bibfnamefont {C.}~\bibnamefont {Godfrin}}, \bibinfo {author} {\bibfnamefont {S.}~\bibnamefont {Kubicek}}, \bibinfo {author} {\bibfnamefont {S.}~\bibnamefont {Massar}}, \bibinfo {author} {\bibfnamefont {Y.}~\bibnamefont {Canvel}}, \bibinfo {author} {\bibfnamefont {J.}~\bibnamefont {Jussot}}, \bibinfo {author} {\bibfnamefont {G.}~\bibnamefont {Simion}}, \bibinfo {author} {\bibfnamefont {M.}~\bibnamefont {Mongillo}}, \bibinfo {author} {\bibfnamefont {D.}~\bibnamefont {Wan}}, \bibinfo {author} {\bibfnamefont {B.}~\bibnamefont {Govoreanu}}, \bibinfo {author} {\bibfnamefont {I.~P.}\ \bibnamefont {Radu}}, \bibinfo {author} {\bibfnamefont {R.}~\bibnamefont {Li}}, \bibinfo {author} {\bibfnamefont {P.}~\bibnamefont {Van~Dorpe}},\ and\ \bibinfo {author} {\bibfnamefont {K.}~\bibnamefont {De~Greve}},\ }\bibfield  {title} {\bibinfo {title} {Low
  charge noise quantum dots with industrial {CMOS} manufacturing},\ }\href {https://doi.org/10.1038/s41534-024-00864-3} {\bibfield  {journal} {\bibinfo  {journal} {npj Quantum Information}\ }\textbf {\bibinfo {volume} {10}},\ \bibinfo {pages} {70} (\bibinfo {year} {2024})}\BibitemShut {NoStop}%
\bibitem [{\citenamefont {Paladino}\ \emph {et~al.}(2014)\citenamefont {Paladino}, \citenamefont {Galperin}, \citenamefont {Falci},\ and\ \citenamefont {Altshuler}}]{paladino_1_2014}%
  \BibitemOpen
  \bibfield  {author} {\bibinfo {author} {\bibfnamefont {E.}~\bibnamefont {Paladino}}, \bibinfo {author} {\bibfnamefont {Y.}~\bibnamefont {Galperin}}, \bibinfo {author} {\bibfnamefont {G.}~\bibnamefont {Falci}},\ and\ \bibinfo {author} {\bibfnamefont {B.}~\bibnamefont {Altshuler}},\ }\bibfield  {title} {\bibinfo {title} {1 / f noise: {Implications} for solid-state quantum information},\ }\href {https://doi.org/10.1103/RevModPhys.86.361} {\bibfield  {journal} {\bibinfo  {journal} {Reviews of Modern Physics}\ }\textbf {\bibinfo {volume} {86}},\ \bibinfo {pages} {361} (\bibinfo {year} {2014})}\BibitemShut {NoStop}%
\bibitem [{\citenamefont {Ferrero}\ \emph {et~al.}(2024)\citenamefont {Ferrero}, \citenamefont {Koch}, \citenamefont {Vogel}, \citenamefont {Schroller}, \citenamefont {Adam}, \citenamefont {Xue}, \citenamefont {Seidler}, \citenamefont {Schreiber}, \citenamefont {Bluhm},\ and\ \citenamefont {Wernsdorfer}}]{ferrero_noise_2024}%
  \BibitemOpen
  \bibfield  {author} {\bibinfo {author} {\bibfnamefont {J.}~\bibnamefont {Ferrero}}, \bibinfo {author} {\bibfnamefont {T.}~\bibnamefont {Koch}}, \bibinfo {author} {\bibfnamefont {S.}~\bibnamefont {Vogel}}, \bibinfo {author} {\bibfnamefont {D.}~\bibnamefont {Schroller}}, \bibinfo {author} {\bibfnamefont {V.}~\bibnamefont {Adam}}, \bibinfo {author} {\bibfnamefont {R.}~\bibnamefont {Xue}}, \bibinfo {author} {\bibfnamefont {I.}~\bibnamefont {Seidler}}, \bibinfo {author} {\bibfnamefont {L.~R.}\ \bibnamefont {Schreiber}}, \bibinfo {author} {\bibfnamefont {H.}~\bibnamefont {Bluhm}},\ and\ \bibinfo {author} {\bibfnamefont {W.}~\bibnamefont {Wernsdorfer}},\ }\bibfield  {title} {\bibinfo {title} {Noise reduction by bias cooling in gated si/sige quantum dots},\ }\href {https://doi.org/10.1063/5.0206632} {\bibfield  {journal} {\bibinfo  {journal} {Applied Physics Letters}\ }\textbf {\bibinfo {volume} {124}},\ \bibinfo {pages} {204002} (\bibinfo {year} {2024})}\BibitemShut {NoStop}%
\bibitem [{\citenamefont {Elzerman}\ \emph {et~al.}(2004)\citenamefont {Elzerman}, \citenamefont {Hanson}, \citenamefont {Willems Van~Beveren}, \citenamefont {Witkamp}, \citenamefont {Vandersypen},\ and\ \citenamefont {Kouwenhoven}}]{elzerman_single-shot_2004}%
  \BibitemOpen
  \bibfield  {author} {\bibinfo {author} {\bibfnamefont {J.~M.}\ \bibnamefont {Elzerman}}, \bibinfo {author} {\bibfnamefont {R.}~\bibnamefont {Hanson}}, \bibinfo {author} {\bibfnamefont {L.~H.}\ \bibnamefont {Willems Van~Beveren}}, \bibinfo {author} {\bibfnamefont {B.}~\bibnamefont {Witkamp}}, \bibinfo {author} {\bibfnamefont {L.~M.~K.}\ \bibnamefont {Vandersypen}},\ and\ \bibinfo {author} {\bibfnamefont {L.~P.}\ \bibnamefont {Kouwenhoven}},\ }\bibfield  {title} {\bibinfo {title} {Single-shot read-out of an individual electron spin in a quantum dot},\ }\href {https://doi.org/10.1038/nature02693} {\bibfield  {journal} {\bibinfo  {journal} {Nature}\ }\textbf {\bibinfo {volume} {430}},\ \bibinfo {pages} {431} (\bibinfo {year} {2004})}\BibitemShut {NoStop}%
\bibitem [{\citenamefont {Huang}\ and\ \citenamefont {Hu}(2014)}]{huang_spin_2014}%
  \BibitemOpen
  \bibfield  {author} {\bibinfo {author} {\bibfnamefont {P.}~\bibnamefont {Huang}}\ and\ \bibinfo {author} {\bibfnamefont {X.}~\bibnamefont {Hu}},\ }\bibfield  {title} {\bibinfo {title} {Spin relaxation in a {Si} quantum dot due to spin-valley mixing},\ }\href {https://doi.org/10.1103/PhysRevB.90.235315} {\bibfield  {journal} {\bibinfo  {journal} {Physical Review B}\ }\textbf {\bibinfo {volume} {90}},\ \bibinfo {pages} {235315} (\bibinfo {year} {2014})}\BibitemShut {NoStop}%
\bibitem [{\citenamefont {Hollmann}\ \emph {et~al.}(2020)\citenamefont {Hollmann}, \citenamefont {Struck}, \citenamefont {Langrock}, \citenamefont {Schmidbauer}, \citenamefont {Schauer}, \citenamefont {Leonhardt}, \citenamefont {Sawano}, \citenamefont {Riemann}, \citenamefont {Abrosimov}, \citenamefont {Bougeard},\ and\ \citenamefont {Schreiber}}]{hollmann_large_2020}%
  \BibitemOpen
  \bibfield  {author} {\bibinfo {author} {\bibfnamefont {A.}~\bibnamefont {Hollmann}}, \bibinfo {author} {\bibfnamefont {T.}~\bibnamefont {Struck}}, \bibinfo {author} {\bibfnamefont {V.}~\bibnamefont {Langrock}}, \bibinfo {author} {\bibfnamefont {A.}~\bibnamefont {Schmidbauer}}, \bibinfo {author} {\bibfnamefont {F.}~\bibnamefont {Schauer}}, \bibinfo {author} {\bibfnamefont {T.}~\bibnamefont {Leonhardt}}, \bibinfo {author} {\bibfnamefont {K.}~\bibnamefont {Sawano}}, \bibinfo {author} {\bibfnamefont {H.}~\bibnamefont {Riemann}}, \bibinfo {author} {\bibfnamefont {N.~V.}\ \bibnamefont {Abrosimov}}, \bibinfo {author} {\bibfnamefont {D.}~\bibnamefont {Bougeard}},\ and\ \bibinfo {author} {\bibfnamefont {L.~R.}\ \bibnamefont {Schreiber}},\ }\bibfield  {title} {\bibinfo {title} {Large, {Tunable} {Valley} {Splitting} and {Single}-{Spin} {Relaxation} {Mechanisms} in a {Si} / {Si} {Ge} {Quantum} {Dot}},\ }\href {https://doi.org/10.1103/PhysRevApplied.13.034068} {\bibfield  {journal} {\bibinfo  {journal} {Physical Review
  Applied}\ }\textbf {\bibinfo {volume} {13}},\ \bibinfo {pages} {034068} (\bibinfo {year} {2020})}\BibitemShut {NoStop}%
\bibitem [{\citenamefont {Borjans}\ \emph {et~al.}(2019)\citenamefont {Borjans}, \citenamefont {Zajac}, \citenamefont {Hazard},\ and\ \citenamefont {Petta}}]{borjans_single-spin_2019}%
  \BibitemOpen
  \bibfield  {author} {\bibinfo {author} {\bibfnamefont {F.}~\bibnamefont {Borjans}}, \bibinfo {author} {\bibfnamefont {D.}~\bibnamefont {Zajac}}, \bibinfo {author} {\bibfnamefont {T.}~\bibnamefont {Hazard}},\ and\ \bibinfo {author} {\bibfnamefont {J.}~\bibnamefont {Petta}},\ }\bibfield  {title} {\bibinfo {title} {Single-{Spin} {Relaxation} in a {Synthetic} {Spin}-{Orbit} {Field}},\ }\href {https://doi.org/10.1103/PhysRevApplied.11.044063} {\bibfield  {journal} {\bibinfo  {journal} {Physical Review Applied}\ }\textbf {\bibinfo {volume} {11}},\ \bibinfo {pages} {044063} (\bibinfo {year} {2019})}\BibitemShut {NoStop}%
\bibitem [{\citenamefont {Burkard}\ \emph {et~al.}(2023)\citenamefont {Burkard}, \citenamefont {Ladd}, \citenamefont {Pan}, \citenamefont {Nichol},\ and\ \citenamefont {Petta}}]{burkard_semiconductor_2023}%
  \BibitemOpen
  \bibfield  {author} {\bibinfo {author} {\bibfnamefont {G.}~\bibnamefont {Burkard}}, \bibinfo {author} {\bibfnamefont {T.~D.}\ \bibnamefont {Ladd}}, \bibinfo {author} {\bibfnamefont {A.}~\bibnamefont {Pan}}, \bibinfo {author} {\bibfnamefont {J.~M.}\ \bibnamefont {Nichol}},\ and\ \bibinfo {author} {\bibfnamefont {J.~R.}\ \bibnamefont {Petta}},\ }\bibfield  {title} {\bibinfo {title} {Semiconductor spin qubits},\ }\href {https://doi.org/10.1103/RevModPhys.95.025003} {\bibfield  {journal} {\bibinfo  {journal} {Reviews of Modern Physics}\ }\textbf {\bibinfo {volume} {95}},\ \bibinfo {pages} {025003} (\bibinfo {year} {2023})}\BibitemShut {NoStop}%
\bibitem [{\citenamefont {Zwanenburg}\ \emph {et~al.}(2013)\citenamefont {Zwanenburg}, \citenamefont {Dzurak}, \citenamefont {Morello}, \citenamefont {Simmons}, \citenamefont {Hollenberg}, \citenamefont {Klimeck}, \citenamefont {Rogge}, \citenamefont {Coppersmith},\ and\ \citenamefont {Eriksson}}]{zwanenburg_silicon_2013}%
  \BibitemOpen
  \bibfield  {author} {\bibinfo {author} {\bibfnamefont {F.~A.}\ \bibnamefont {Zwanenburg}}, \bibinfo {author} {\bibfnamefont {A.~S.}\ \bibnamefont {Dzurak}}, \bibinfo {author} {\bibfnamefont {A.}~\bibnamefont {Morello}}, \bibinfo {author} {\bibfnamefont {M.~Y.}\ \bibnamefont {Simmons}}, \bibinfo {author} {\bibfnamefont {L.~C.~L.}\ \bibnamefont {Hollenberg}}, \bibinfo {author} {\bibfnamefont {G.}~\bibnamefont {Klimeck}}, \bibinfo {author} {\bibfnamefont {S.}~\bibnamefont {Rogge}}, \bibinfo {author} {\bibfnamefont {S.~N.}\ \bibnamefont {Coppersmith}},\ and\ \bibinfo {author} {\bibfnamefont {M.~A.}\ \bibnamefont {Eriksson}},\ }\bibfield  {title} {\bibinfo {title} {Silicon quantum electronics},\ }\href {https://doi.org/10.1103/RevModPhys.85.961} {\bibfield  {journal} {\bibinfo  {journal} {Reviews of Modern Physics}\ }\textbf {\bibinfo {volume} {85}},\ \bibinfo {pages} {961} (\bibinfo {year} {2013})}\BibitemShut {NoStop}%
\bibitem [{\citenamefont {Hosseinkhani}\ and\ \citenamefont {Burkard}(2020)}]{hosseinkhani_electromagnetic_2020}%
  \BibitemOpen
  \bibfield  {author} {\bibinfo {author} {\bibfnamefont {A.}~\bibnamefont {Hosseinkhani}}\ and\ \bibinfo {author} {\bibfnamefont {G.}~\bibnamefont {Burkard}},\ }\bibfield  {title} {\bibinfo {title} {Electromagnetic control of valley splitting in ideal and disordered {Si} quantum dots},\ }\href {https://doi.org/10.1103/PhysRevResearch.2.043180} {\bibfield  {journal} {\bibinfo  {journal} {Physical Review Research}\ }\textbf {\bibinfo {volume} {2}},\ \bibinfo {pages} {043180} (\bibinfo {year} {2020})}\BibitemShut {NoStop}%
\bibitem [{\citenamefont {Volmer}\ \emph {et~al.}(2024)\citenamefont {Volmer}, \citenamefont {Struck}, \citenamefont {Sala}, \citenamefont {Chen}, \citenamefont {Oberländer}, \citenamefont {Offermann}, \citenamefont {Xue}, \citenamefont {Visser}, \citenamefont {Tu}, \citenamefont {Trellenkamp}, \citenamefont {Cywinski}, \citenamefont {Bluhm},\ and\ \citenamefont {Schreiber}}]{volmer_mapping_2024}%
  \BibitemOpen
  \bibfield  {author} {\bibinfo {author} {\bibfnamefont {M.}~\bibnamefont {Volmer}}, \bibinfo {author} {\bibfnamefont {T.}~\bibnamefont {Struck}}, \bibinfo {author} {\bibfnamefont {A.}~\bibnamefont {Sala}}, \bibinfo {author} {\bibfnamefont {B.}~\bibnamefont {Chen}}, \bibinfo {author} {\bibfnamefont {M.}~\bibnamefont {Oberländer}}, \bibinfo {author} {\bibfnamefont {T.}~\bibnamefont {Offermann}}, \bibinfo {author} {\bibfnamefont {R.}~\bibnamefont {Xue}}, \bibinfo {author} {\bibfnamefont {L.}~\bibnamefont {Visser}}, \bibinfo {author} {\bibfnamefont {J.-S.}\ \bibnamefont {Tu}}, \bibinfo {author} {\bibfnamefont {S.}~\bibnamefont {Trellenkamp}}, \bibinfo {author} {\bibfnamefont {L.}~\bibnamefont {Cywinski}}, \bibinfo {author} {\bibfnamefont {H.}~\bibnamefont {Bluhm}},\ and\ \bibinfo {author} {\bibfnamefont {L.~R.}\ \bibnamefont {Schreiber}},\ }\bibfield  {title} {\bibinfo {title} {Mapping of valley splitting by conveyor-mode spin-coherent electron shuttling},\ }\href {https://doi.org/10.1038/s41534-024-00852-7}
  {\bibfield  {journal} {\bibinfo  {journal} {npj Quantum Information}\ }\textbf {\bibinfo {volume} {10}},\ \bibinfo {pages} {61} (\bibinfo {year} {2024})}\BibitemShut {NoStop}%
\bibitem [{\citenamefont {Yang}\ \emph {et~al.}(2013)\citenamefont {Yang}, \citenamefont {Rossi}, \citenamefont {Ruskov}, \citenamefont {Lai}, \citenamefont {Mohiyaddin}, \citenamefont {Lee}, \citenamefont {Tahan}, \citenamefont {Klimeck}, \citenamefont {Morello},\ and\ \citenamefont {Dzurak}}]{yang_spin-valley_2013}%
  \BibitemOpen
  \bibfield  {author} {\bibinfo {author} {\bibfnamefont {C.~H.}\ \bibnamefont {Yang}}, \bibinfo {author} {\bibfnamefont {A.}~\bibnamefont {Rossi}}, \bibinfo {author} {\bibfnamefont {R.}~\bibnamefont {Ruskov}}, \bibinfo {author} {\bibfnamefont {N.~S.}\ \bibnamefont {Lai}}, \bibinfo {author} {\bibfnamefont {F.~A.}\ \bibnamefont {Mohiyaddin}}, \bibinfo {author} {\bibfnamefont {S.}~\bibnamefont {Lee}}, \bibinfo {author} {\bibfnamefont {C.}~\bibnamefont {Tahan}}, \bibinfo {author} {\bibfnamefont {G.}~\bibnamefont {Klimeck}}, \bibinfo {author} {\bibfnamefont {A.}~\bibnamefont {Morello}},\ and\ \bibinfo {author} {\bibfnamefont {A.~S.}\ \bibnamefont {Dzurak}},\ }\bibfield  {title} {\bibinfo {title} {Spin-valley lifetimes in a silicon quantum dot with tunable valley splitting},\ }\href {https://doi.org/10.1038/ncomms3069} {\bibfield  {journal} {\bibinfo  {journal} {Nature Communications}\ }\textbf {\bibinfo {volume} {4}},\ \bibinfo {pages} {2069} (\bibinfo {year} {2013})}\BibitemShut {NoStop}%
\bibitem [{\citenamefont {Petit}\ \emph {et~al.}(2018)\citenamefont {Petit}, \citenamefont {Boter}, \citenamefont {Eenink}, \citenamefont {Droulers}, \citenamefont {Tagliaferri}, \citenamefont {Li}, \citenamefont {Franke}, \citenamefont {Singh}, \citenamefont {Clarke}, \citenamefont {Schouten}, \citenamefont {Dobrovitski}, \citenamefont {Vandersypen},\ and\ \citenamefont {Veldhorst}}]{petit_spin_2018}%
  \BibitemOpen
  \bibfield  {author} {\bibinfo {author} {\bibfnamefont {L.}~\bibnamefont {Petit}}, \bibinfo {author} {\bibfnamefont {J.}~\bibnamefont {Boter}}, \bibinfo {author} {\bibfnamefont {H.}~\bibnamefont {Eenink}}, \bibinfo {author} {\bibfnamefont {G.}~\bibnamefont {Droulers}}, \bibinfo {author} {\bibfnamefont {M.}~\bibnamefont {Tagliaferri}}, \bibinfo {author} {\bibfnamefont {R.}~\bibnamefont {Li}}, \bibinfo {author} {\bibfnamefont {D.}~\bibnamefont {Franke}}, \bibinfo {author} {\bibfnamefont {K.}~\bibnamefont {Singh}}, \bibinfo {author} {\bibfnamefont {J.}~\bibnamefont {Clarke}}, \bibinfo {author} {\bibfnamefont {R.}~\bibnamefont {Schouten}}, \bibinfo {author} {\bibfnamefont {V.}~\bibnamefont {Dobrovitski}}, \bibinfo {author} {\bibfnamefont {L.}~\bibnamefont {Vandersypen}},\ and\ \bibinfo {author} {\bibfnamefont {M.}~\bibnamefont {Veldhorst}},\ }\bibfield  {title} {\bibinfo {title} {Spin {Lifetime} and {Charge} {Noise} in {Hot} {Silicon} {Quantum} {Dot} {Qubits}},\ }\href
  {https://doi.org/10.1103/PhysRevLett.121.076801} {\bibfield  {journal} {\bibinfo  {journal} {Physical Review Letters}\ }\textbf {\bibinfo {volume} {121}},\ \bibinfo {pages} {076801} (\bibinfo {year} {2018})}\BibitemShut {NoStop}%
\bibitem [{\citenamefont {Stano}\ and\ \citenamefont {Loss}(2022)}]{stano_review_2022}%
  \BibitemOpen
  \bibfield  {author} {\bibinfo {author} {\bibfnamefont {P.}~\bibnamefont {Stano}}\ and\ \bibinfo {author} {\bibfnamefont {D.}~\bibnamefont {Loss}},\ }\bibfield  {title} {\bibinfo {title} {Review of performance metrics of spin qubits in gated semiconducting nanostructures},\ }\href {https://doi.org/10.1038/s42254-022-00484-w} {\bibfield  {journal} {\bibinfo  {journal} {Nature Reviews Physics}\ }\textbf {\bibinfo {volume} {4}},\ \bibinfo {pages} {672} (\bibinfo {year} {2022})}\BibitemShut {NoStop}%
\bibitem [{\citenamefont {Hanson}\ \emph {et~al.}(2007)\citenamefont {Hanson}, \citenamefont {Kouwenhoven}, \citenamefont {Petta}, \citenamefont {Tarucha},\ and\ \citenamefont {Vandersypen}}]{hanson_spins_2007}%
  \BibitemOpen
  \bibfield  {author} {\bibinfo {author} {\bibfnamefont {R.}~\bibnamefont {Hanson}}, \bibinfo {author} {\bibfnamefont {L.~P.}\ \bibnamefont {Kouwenhoven}}, \bibinfo {author} {\bibfnamefont {J.~R.}\ \bibnamefont {Petta}}, \bibinfo {author} {\bibfnamefont {S.}~\bibnamefont {Tarucha}},\ and\ \bibinfo {author} {\bibfnamefont {L.~M.~K.}\ \bibnamefont {Vandersypen}},\ }\bibfield  {title} {\bibinfo {title} {Spins in few-electron quantum dots},\ }\href {https://doi.org/10.1103/RevModPhys.79.1217} {\bibfield  {journal} {\bibinfo  {journal} {Reviews of Modern Physics}\ }\textbf {\bibinfo {volume} {79}},\ \bibinfo {pages} {1217} (\bibinfo {year} {2007})}\BibitemShut {NoStop}%
\bibitem [{\citenamefont {McCourt}\ \emph {et~al.}(2023)\citenamefont {McCourt}, \citenamefont {Neill}, \citenamefont {Lee}, \citenamefont {Quintana}, \citenamefont {Chen}, \citenamefont {Kelly}, \citenamefont {Marshall}, \citenamefont {Smelyanskiy}, \citenamefont {Dykman}, \citenamefont {Korotkov}, \citenamefont {Chuang},\ and\ \citenamefont {Petukhov}}]{mccourt_learning_2023}%
  \BibitemOpen
  \bibfield  {author} {\bibinfo {author} {\bibfnamefont {T.}~\bibnamefont {McCourt}}, \bibinfo {author} {\bibfnamefont {C.}~\bibnamefont {Neill}}, \bibinfo {author} {\bibfnamefont {K.}~\bibnamefont {Lee}}, \bibinfo {author} {\bibfnamefont {C.}~\bibnamefont {Quintana}}, \bibinfo {author} {\bibfnamefont {Y.}~\bibnamefont {Chen}}, \bibinfo {author} {\bibfnamefont {J.}~\bibnamefont {Kelly}}, \bibinfo {author} {\bibfnamefont {J.}~\bibnamefont {Marshall}}, \bibinfo {author} {\bibfnamefont {V.~N.}\ \bibnamefont {Smelyanskiy}}, \bibinfo {author} {\bibfnamefont {M.~I.}\ \bibnamefont {Dykman}}, \bibinfo {author} {\bibfnamefont {A.}~\bibnamefont {Korotkov}}, \bibinfo {author} {\bibfnamefont {I.~L.}\ \bibnamefont {Chuang}},\ and\ \bibinfo {author} {\bibfnamefont {A.~G.}\ \bibnamefont {Petukhov}},\ }\bibfield  {title} {\bibinfo {title} {Learning noise via dynamical decoupling of entangled qubits},\ }\href {https://doi.org/10.1103/PhysRevA.107.052610} {\bibfield  {journal} {\bibinfo  {journal} {Physical Review A}\ }\textbf
  {\bibinfo {volume} {107}},\ \bibinfo {pages} {052610} (\bibinfo {year} {2023})}\BibitemShut {NoStop}%
\bibitem [{\citenamefont {Muhonen}\ \emph {et~al.}(2014)\citenamefont {Muhonen}, \citenamefont {Dehollain}, \citenamefont {Laucht}, \citenamefont {Hudson}, \citenamefont {Kalra}, \citenamefont {Sekiguchi}, \citenamefont {Itoh}, \citenamefont {Jamieson}, \citenamefont {McCallum}, \citenamefont {Dzurak},\ and\ \citenamefont {Morello}}]{muhonen_storing_2014}%
  \BibitemOpen
  \bibfield  {author} {\bibinfo {author} {\bibfnamefont {J.~T.}\ \bibnamefont {Muhonen}}, \bibinfo {author} {\bibfnamefont {J.~P.}\ \bibnamefont {Dehollain}}, \bibinfo {author} {\bibfnamefont {A.}~\bibnamefont {Laucht}}, \bibinfo {author} {\bibfnamefont {F.~E.}\ \bibnamefont {Hudson}}, \bibinfo {author} {\bibfnamefont {R.}~\bibnamefont {Kalra}}, \bibinfo {author} {\bibfnamefont {T.}~\bibnamefont {Sekiguchi}}, \bibinfo {author} {\bibfnamefont {K.~M.}\ \bibnamefont {Itoh}}, \bibinfo {author} {\bibfnamefont {D.~N.}\ \bibnamefont {Jamieson}}, \bibinfo {author} {\bibfnamefont {J.~C.}\ \bibnamefont {McCallum}}, \bibinfo {author} {\bibfnamefont {A.~S.}\ \bibnamefont {Dzurak}},\ and\ \bibinfo {author} {\bibfnamefont {A.}~\bibnamefont {Morello}},\ }\bibfield  {title} {\bibinfo {title} {Storing quantum information for 30 seconds in a nanoelectronic device},\ }\href {https://doi.org/10.1038/nnano.2014.211} {\bibfield  {journal} {\bibinfo  {journal} {Nature Nanotechnology}\ }\textbf {\bibinfo {volume} {9}},\ \bibinfo
  {pages} {986} (\bibinfo {year} {2014})}\BibitemShut {NoStop}%
\bibitem [{\citenamefont {Welch}(1967)}]{welch_use_1967}%
  \BibitemOpen
  \bibfield  {author} {\bibinfo {author} {\bibfnamefont {P.}~\bibnamefont {Welch}},\ }\bibfield  {title} {\bibinfo {title} {The use of fast {Fourier} transform for the estimation of power spectra: {A} method based on time averaging over short, modified periodograms},\ }\href {https://doi.org/10.1109/TAU.1967.1161901} {\bibfield  {journal} {\bibinfo  {journal} {IEEE Transactions on Audio and Electroacoustics}\ }\textbf {\bibinfo {volume} {15}},\ \bibinfo {pages} {70} (\bibinfo {year} {1967})}\BibitemShut {NoStop}%
\bibitem [{\citenamefont {Epstein}\ \emph {et~al.}(2014)\citenamefont {Epstein}, \citenamefont {Cross}, \citenamefont {Magesan},\ and\ \citenamefont {Gambetta}}]{epstein_investigating_2014}%
  \BibitemOpen
  \bibfield  {author} {\bibinfo {author} {\bibfnamefont {J.~M.}\ \bibnamefont {Epstein}}, \bibinfo {author} {\bibfnamefont {A.~W.}\ \bibnamefont {Cross}}, \bibinfo {author} {\bibfnamefont {E.}~\bibnamefont {Magesan}},\ and\ \bibinfo {author} {\bibfnamefont {J.~M.}\ \bibnamefont {Gambetta}},\ }\bibfield  {title} {\bibinfo {title} {Investigating the limits of randomized benchmarking protocols},\ }\href {https://doi.org/10.1103/PhysRevA.89.062321} {\bibfield  {journal} {\bibinfo  {journal} {Physical Review A}\ }\textbf {\bibinfo {volume} {89}},\ \bibinfo {pages} {062321} (\bibinfo {year} {2014})}\BibitemShut {NoStop}%
\bibitem [{\citenamefont {Dumoulin~Stuyck}\ \emph {et~al.}(2021)\citenamefont {Dumoulin~Stuyck}, \citenamefont {Mohiyaddin}, \citenamefont {Li}, \citenamefont {Heyns}, \citenamefont {Govoreanu},\ and\ \citenamefont {Radu}}]{dumoulin_stuyck_low_2021}%
  \BibitemOpen
  \bibfield  {author} {\bibinfo {author} {\bibfnamefont {N.~I.}\ \bibnamefont {Dumoulin~Stuyck}}, \bibinfo {author} {\bibfnamefont {F.~A.}\ \bibnamefont {Mohiyaddin}}, \bibinfo {author} {\bibfnamefont {R.}~\bibnamefont {Li}}, \bibinfo {author} {\bibfnamefont {M.}~\bibnamefont {Heyns}}, \bibinfo {author} {\bibfnamefont {B.}~\bibnamefont {Govoreanu}},\ and\ \bibinfo {author} {\bibfnamefont {I.~P.}\ \bibnamefont {Radu}},\ }\bibfield  {title} {\bibinfo {title} {Low dephasing and robust micromagnet designs for silicon spin qubits},\ }\href {https://doi.org/10.1063/5.0059939} {\bibfield  {journal} {\bibinfo  {journal} {Applied Physics Letters}\ }\textbf {\bibinfo {volume} {119}},\ \bibinfo {pages} {094001} (\bibinfo {year} {2021})}\BibitemShut {NoStop}%
\end{thebibliography}%

\end{document}